\documentclass[onecolumn,amsmath,amssymb,12pt,superscriptaddress,nofootinbib]{revtex4}
\pdfoutput=1

\usepackage[latin1]{inputenc}
\usepackage[english]{babel}
\usepackage{amssymb}
\usepackage{amsmath}
\usepackage{amsthm}
\usepackage[]{graphicx}
\usepackage[]{subfigure}
\usepackage{tensor}
\usepackage{color}
\usepackage{cancel}
\usepackage{setspace}
\usepackage{fancyhdr}
\usepackage{framed}
\usepackage{tikz}
\usepackage[bookmarks,linktocpage, colorlinks=true, plainpages = false, citecolor = blue,  linkcolor=blue, urlcolor = blue, filecolor = blue]{hyperref} 

\begin{document}

\allowdisplaybreaks
\begin{titlepage}

\title{Lorentzian Quantum Cosmology \vspace{.3in}}

\author{Job Feldbrugge}
\email{jfeldbrugge@perimeterinstitute.ca}
\affiliation{Perimeter Institute, 31 Caroline St N, Ontario, Canada}
\author{Jean-Luc Lehners}
\email{jlehners@aei.mpg.de}
\affiliation{Max--Planck--Institute for Gravitational Physics (Albert--Einstein--Institute), 14476 Potsdam, Germany}
\author{Neil Turok}
\email{nturok@perimeterinstitute.ca}
\affiliation{Perimeter Institute, 31 Caroline St N, Ontario, Canada}

\begin{abstract}
\vspace{.3in} \noindent 
We argue that the Lorentzian path integral is a better starting point for quantum cosmology than the Euclidean version. In particular, we revisit the mini-superspace calculation of the Feynman path integral for quantum gravity with a positive cosmological constant. Instead of rotating to Euclidean time, we deform the contour of integration over metrics into the complex plane,  exploiting Picard-Lefschetz theory to transform the path integral from a conditionally convergent integral into an absolutely convergent one. We show that this procedure unambiguously determines which semiclassical saddle point solutions are relevant to the quantum mechanical amplitude. Imposing ``no-boundary'' initial conditions, {\it i.e.}, restricting attention to regular, complex metrics with no initial boundary, we find that the dominant saddle contributes a semiclassical exponential factor which is precisely the {\it inverse} of the famous Hartle-Hawking result. 
\end{abstract}
\maketitle

\end{titlepage}

\tableofcontents


\section{Introduction}
Any theory of cosmology must provide both a successful description of the dynamics and an explanation of the initial state. Proposals for the initial quantum state include the ``no-boundary'' proposal of Hartle and Hawing \cite{Hawking:1981gb,Hartle:1983ai,Hawking:1983hj,Hartle:2008ng} and the tunneling wavefunction of Vilenkin \cite{Vilenkin:1982de,Vilenkin:1983xq,Vilenkin:1984wp,Vilenkin:1986cy}. 

The ``no-boundary'' proposal is usually framed in terms of the path integral for the Euclidean (or Riemannian) version of general relativity. 
One motivation provided was that the Euclidean path integral would have nicer convergence properties, making it better defined, in analogy with Euclidean (Wick-rotated) quantum field theory~\cite{Hawking:1981gb}. Another was that the Euclidean approach would straightforwardly allow for the inclusion of topologically non-trivial manifolds~\cite{Hawking:1981gb}. However, the Euclidean approach to quantum gravity was only partly successful. An immediate problem is that the Euclidean action is unbounded below~\cite{Gibbons:1977zz}. One well-known aspect of this is the conformal factor problem; gradients in the overall scale factor of the metric contribute negatively to the Euclidean action and typically render it unbounded below. Another aspect stems precisely from the inclusion of topologically non-trivial manifolds, whose action can be shown to be unbounded both above and below~\cite{Gibbons:1977zz}. 
Unfortunately, this means that specifying a boundary condition, such as the ``no-boundary'' condition, is insufficient to uniquely define the Euclidean path integral. Additional input is required to determine the complex contour over which one should integrate Euclidean metrics~\cite{Gibbons:1978ac}. In the absence of such a prescription, the Euclidean path integral for quantum gravity is ill-defined at best.  

An alternate approach, followed in much of the literature, is based on solving the homogeneous Wheeler-DeWitt equation for the ``wavefunction of the universe.''  However, there are also significant ambiguities in this approach, since there is in principle an infinite degree of freedom in the choice of boundary conditions on superspace. There are both real and complex solutions and different authors have taken different views about which of these should be taken into account. The tunneling wavefunction, for example, was developed as a particular choice for a complex solution, representing an outgoing, expanding universe only, while the ``no-boundary'' proposal was presented as a real solution representing a quantum superposition of an expanding universe with a collapsing one. 

More recently, a holographic approach to quantum cosmology, based on the AdS/CFT correspondence, has also been advocated~\cite{Hertog:2004rz,Hertog:2005hu}. The problem with this approach is that it requires boundary conditions which strongly influence the dynamics in the bulk. If the boundary conditions are AdS-invariant, the cosmological dynamics (including the Jeans instability on large scales) is eliminated altogether. More general boundary conditions typically either allow instabilities which are difficult to resolve~\cite{Harlow:2010az}, or they affect the bulk dynamics so strongly that any connection to four-dimensional general relativistic cosmology is unclear~\cite{Maldacena:2010un}. More general holographic scenarios have also been proposed~\cite{McFadden:2009fg}, based on analytically continuing Newton's constant and other fundamental parameters in order to obtain a well-defined dual 3d theory. While this procedure works formally for certain specific cosmological backgrounds, it does not so far seem to address more fundamental questions such as the likelihood of such backgrounds. 

In the present paper we argue for what seems to us a much simpler, clearer and more general approach. We start from the Lorentzian path integral, defined as a functional integral over real Lorentzian metrics. This has several immediate advantages over the Euclidean path integral. There is no conformal factor problem. The Lorentzian path integral naturally incorporates notions of causality and unitarity,  as well as boundary conditions specified in terms of the initial and final three-geometry, eliminating the corresponding ambiguities in the Wheeler-DeWitt wavefunction approach. The only disadvantage of the Lorentzian path integral is a technical, although very important one: it is a highly oscillatory integral, whose convergence is not  obvious. In fact, from its definition as an infinite-dimensional integral over an infinite measure of the phase $e^{i S/\hbar}$, it is clearly {\it not} absolutely convergent. In simple examples, however, we find that it is {\it conditionally} convergent, for very simple reasons which are likely to extend to the general case.  Conditionally convergent integrals dependent on the order of ``summation,'' just as conditionally convergent series depend on the ordering of the sum, and for this reason they are more subtle to treat. A classic example is the Dirichlet integral $\int_{-\infty}^\infty \frac{\sin(x)}{x}\mathrm{d}x = \pi$. 

Traditionally, convergence of the path integral in quantum field theory has been ensured either with Feynman's $i\epsilon$ regularization or, more rigorously, via a Wick rotation to imaginary time, the procedure which motivated the Euclidean path integral approach to quantum gravity. However, in flat space quantum field theory, Wick rotation is much more natural. It exploits the global symmetries of Minkowski spacetime to convert the Lorentz group into a compact rotation group. There are no such symmetries in a generic spacetime. Furthermore, in general relativity, the time coordinate is an arbitrary, unphysical quantity, so analytically continuing it into the complex plane, with no control over its range, has no comparable justification. 

Picard-Lefschetz theory provides an alternative procedure to convert conditionally convergent integrals into absolutely convergent integrals. The idea is to deform the contour of integration from the real axis into the complex plane in such a way as to make the integral absolutely convergent. In the context of a theory in which spacetime itself is dynamical, this seems far more natural than trying to generalize the Wick rotation. One deforms the integral over physical quantities, {\it i.e.}, the four-geometries, which are being integrated over, into the complex plane, while holding the  three-geometry boundaries fixed. Formally, one can appeal to an infinite dimensional version of Cauchy's theorem to ensure that the result is equal to the original Lorentzian path integral.  An elementary example of the Picard-Lefschetz approach was given in \cite{Gielen:2016fdb}, where the Feynman propagator for a relativistic particle was derived from the path integral without any need for the $i\epsilon$ prescription or a Wick rotation. Here, we shall apply the same approach to the minisuperspace path integral for quantum gravity. This gives a well-defined answer while properly incorporating causality and the correct boundary conditions. As we shall see, Picard-Lefschetz theory neatly and unambiguously determines the combination of classical saddle point solutions contributing to semiclassical quantum gravity amplitudes. We have to admit we are puzzled as to why this method, which with hindsight seems by far the most natural and obvious one, has not, as far as we know, been previously advanced in the context of quantum cosmology. 

We shall illustrate our approach in the simplest of toy cosmological models, namely a homogeneous, isotropic, closed universe with a cosmological constant $\Lambda$. We shall compute the quantum mechanical propagator with various boundary conditions, namely classical, non-classical and ``no-boundary'' boundary conditions. Our results for the ``no boundary'' case differ from those of Hartle and Hawking, for an easily understandable reason.

The  Lorentzian path integral is defined as an integral over a phase $e^{i S/\hbar}$, with the action $S$ a real function of real dynamical variables. If one deforms the integration contour into the complex plane for these variables, running through a complex saddle point, as we shall show, one necessarily does so by ``sliding down'' a contour of steepest ascent from the saddle point which intersects the real axis. The real part of the exponent ${\rm Re}[iS_{cl}/\hbar] $, which determines the semiclassical factor in the quantum mechanical amplitude, necessarily decreases on the way down. Since the real part of the semiclassical exponent starts out zero on the real axis, {\it it must be negative at any relevant saddle point}. Such semiclassical factors, by this argument, can only suppress, and never enhance, a quantum mechanical amplitude. 

As is well known, Euclidean quantum gravity yields a {\it positive} real part for the semiclassical exponent, in the case of our simple cosmology. The classical saddle point solution is just a four-sphere, continued at its equator to de Sitter spacetime. In units where $8 \pi G=1$, it yields a semiclassical factor $|e^{i S_{cl}/\hbar}|=e^{{\rm Re}[iS_{cl}/\hbar]}=e^{+12\pi^2/(\hbar \Lambda)}$. This is clearly inconsistent with our argument, so we can safely conclude that the Euclidean solution is not relevant to the Lorentzian path integral. 

Instead, we find that there is a different classical solution, contributing precisely the inverse semiclassical factor, {\it i.e.} a suppression. The reason is simply that the equations of motion are real. If a complex solution exists, its complex conjugate must also be a solution. But the complex conjugate solution has the complex conjugate value for the classical action, so the real part of the semiclassical exponent, Re$[i S_{cl}/\hbar]$, has the opposite sign. This complex conjugate saddle point therefore can be (and, we shall show, is) relevant to the Lorentzian path integral, and gives a semiclassical exponential factor of $e^{-12\pi^2/(\hbar \Lambda)}$, precisely the inverse of the Hartle-Hawking result.  This is the crux of our argument, which the remainder of the paper is devoted to fleshing out in detail.

The semiclassical factor we obtain agrees with Vilenkin's ``tunneling'' proposal, for this simple cosmology. Since the logic is quite different, however, it remains to be seen whether the correspondence persists for more general models. Note also that we are not employing an ``inverse'' Wick rotation, advocated by Linde \cite{Linde:1983mx}. That prescription is well known to be problematic since it leads to a divergent measure for the perturbations. Since we are always considering the Lorentzian path integral, we never perform a Wick rotation. The appropriate contours for the path integral are completely specified by requiring that a) they are continously deformable to contours running over real, Lorentzian spacetime metrics and b) they follow steepest descent contours, along which the path integral is absolutely convergent. These criteria are clearly the appropriate ones for considering semiclassical, Lorentzian amplitudes in general relativity: in our view, there are no good reasons for adopting (and many good reasons {\it not} to adopt) different criteria in quantum cosmology. 

As our argument above already indicates, in these toy cosmologies (and, most likely, in general) the path integral over real Lorentzian metrics {\it cannot} be deformed to a Euclidean contour. Just to be clear, there {\it is} a saddle point of the Euclidean action -- Hartle and Hawking's solution.  And there {\it is} an integration contour running through this saddle point which gives a convergent result -- the steepest descent contour through this saddle. However, this contour bears no relation either to the Lorentzian path integral, {\it or} to one taken over Euclidean metrics, which is a meaningless divergent integral. Instead, the steepest descent contour through the Euclidean (Hartle-Hawking) solution defines an intrinsically complex theory bearing no relation to quantum mechanics or the Lorentzian path integral, and from which, we claim, there is no reason to expect causal or unitary behavior to emerge. 

In summary, the Lorentzian approach we take has several manifest advantages: (i) it starts from a theory with clear notions of causality and quantum-mechanical unitarity, (ii) it does not suffer from a conformal factor problem and (iii) with the Picard-Lefschetz approach to ensuring absolute convergence, it has a chance of being mathematically well defined. We emphasize that few of our detailed calculations in this paper are new or original.  For the most part they recapitulate the analyses of earlier authors, some performed decades ago. Our sole claim to originality is to demonstrate that very minimal and well-founded principles can and do substantially clarify the rules of the game.

This paper is organised as follows. In section \ref{sec:PL} we provide a simple introduction to Picard-Lefschetz theory. In section \ref{section:HH}, we apply this method to the mini-superspace path integral for general relativity with a cosmological constant, for various boundary conditions. In particular, we check cases where the boundary conditions lead to purely classical evolution. Subsequently, we analyze the ``no-boundary'' initial condition in detail, as the main focus of this paper. For completeness we also consider the case with non-classical boundary conditions, where both the initial and final scale factor are smaller than the waist of the de Sitter hyperboloid. In section \ref{sectionWdW} we relate the path integral description to the canonical formalism and the Wheeler-DeWitt equation.  We summarize our findings, and point to future research directions, in section \ref{sectionConclusions}.


\section{Picard-Lefschetz approach to oscillatory integrals} \label{sec:PL}

Picard-Lefschetz theory deals with oscillatory integrals like
\begin{align}
\label{eq:integral}
I = \int_D \mathrm{d}x\, e^{iS[x]/\hbar},
\end{align}
where $\hbar$ is a real parameter, the action $S[x]$ is a real-valued function and the integral is taken over a real domain $D$, usually defined by the singularities of the integrand or, in higher dimensional or path integral cases, its partial integrals.  One is typically interested in the behavior of the integral for small values of the parameter $\hbar$: in quantum mechanical applications, taking $\hbar$ to zero is a nice way to study the classical limit. Picard-Lefschetz theory was originally developed and applied to ordinary integrals, in finite dimension, for example, in the work of Arnol'd et.\ al.\ \cite{Arnold:singularities}. More recently, Witten~\cite{Witten:2010cx} and others have discussed its use in quantum mechanical path integrals.  For example, it has been used to develop new Monte-Carlo techniques capable of addressing the notorious ``sign problem'' in some quantum field theories~\cite{Cristoforetti:2013qaa}. 

In the main part of this paper, we have a far humbler goal. We address simple minisuperspace models of quantum cosmology, which reduce to a single one-dimensional integral. Hence, for the purposes of this brief introduction, we shall review the theory in its most trivial case. It is important to note, however, that in principle Picard-Lefschetz theory may equally be applied in higher dimensions and even, in principle, in the infinite-dimensional context relevant to physically realistic path integrals. 

When faced with an integral in the form of (\ref{eq:integral}), the idea of Picard-Lefshetz theory is to interpret $S[x]$ as a holomorphic function of $x \in \mathbb{C},$ the complex plane. Cauchy's theorem allows us to deform the  integration contour from the real domain $D$ on the real $x$-axis into a contour we now call $\mathcal{C}$ in the complex $x$-plane,  while keeping its endpoints fixed. In particular, we seek to deform $\mathcal{C}$ into a ``steepest descent'' contour passing through one or more critical points of $S[x]$, {\it i.e.} points where $\partial_x S =0$. By the Cauchy-Riemann equations, the real part of the exponent, Re$[i S[x]]$, which controls the magnitude of the integrand, has a saddle point in the real two-dimensional $($Re$[x],$Im$[x])$-plane there. The steepest descent contour through the saddle point is defined as the path along which Re$[i S[x]]$ decreases as rapidly as possible.  

A simple example is provided by $S[x] =x^2$, with a critical point at $x=0$. Writing $x=$Re$[x]+i \,$Im$[x]$, we have Re$[iS[x]]=-2$ Re$[x]$Im$[x]$. The magnitude of the integrand decreases most rapidly along the contour Im$[x]$=+Re$[x]$ which is the steepest descent contour. Conversely, it increases most rapidly along the contour Im$[x]=-$Re$[x]$, which is the steepest ascent contour. As we shall discuss, steepest descent contours generically lead to convergent integrals, and in this case they are known as Lefschetz thimbles ${\cal J}_\sigma$.

In more detail, we write the exponent $\mathcal{I}=iS/\hbar$ and its argument $x$ in terms of their real and imaginary parts, $\mathcal{I}=h+i H$ and $x=u^1+iu^2$.  Downward flow is then defined by 
\begin{equation}
\frac{\mathrm{d}u^i}{\mathrm{d}\lambda} = -g^{ij}\frac{\partial h}{\partial u^j}\,,
\label{eq:dw}
\end{equation}
with $\lambda$ a parameter along the flow and $g_{ij}$ a Riemannian metric introduced on the complex plane. The real part of the exponent $h$  (known as the Morse function) decreases on such a flow away from its critical points, because $\frac{\mathrm{d}h}{\mathrm{d} \lambda} = \sum_i\frac{\partial h}{\partial u^i}\frac{\mathrm{d}u^i}{\mathrm{d}\lambda} = -\sum_i\left(\frac{\partial h}{\partial u^i}\right)^2<0$, with the fastest rate of decrease occuring in the direction of ``steepest descent'', which maximises the magnitude of the gradient. Defining the latter requires that we introduce a metric. Witten points out that the freedom to choose this metric may be exploited in interesting ways~\cite{Witten:2010cx}. 

\begin{figure}[h] 
\begin{minipage}{0.5\textwidth}
		\includegraphics[width=0.7\textwidth]{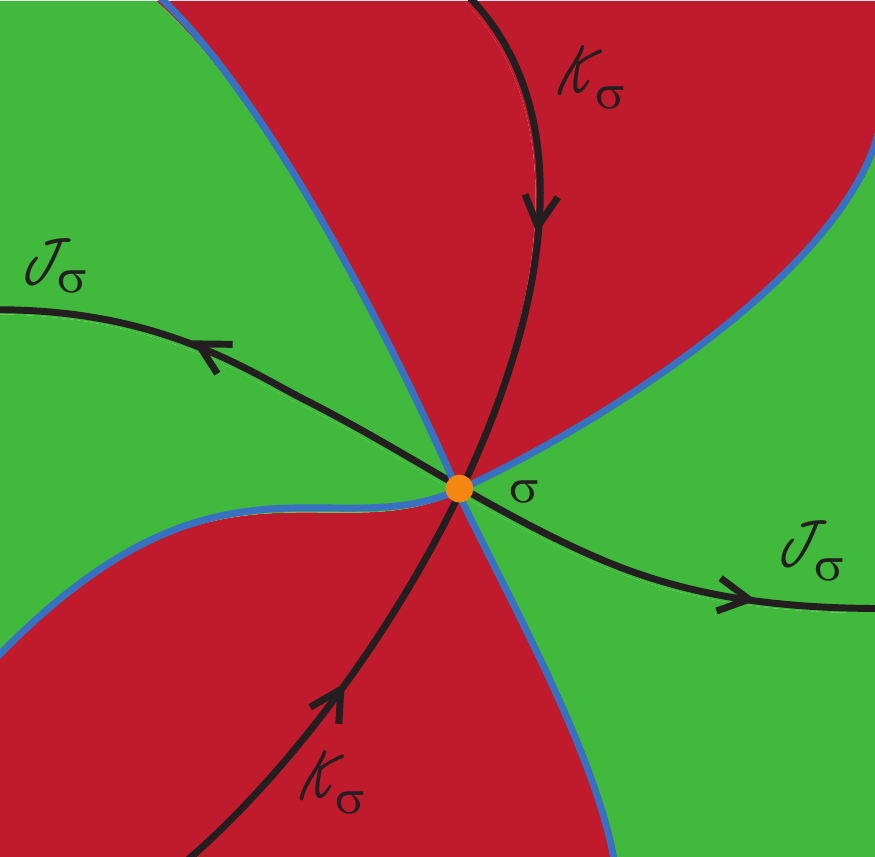}
	\end{minipage}%
	\begin{minipage}{0.5\textwidth}
		\includegraphics[width=0.95\textwidth]{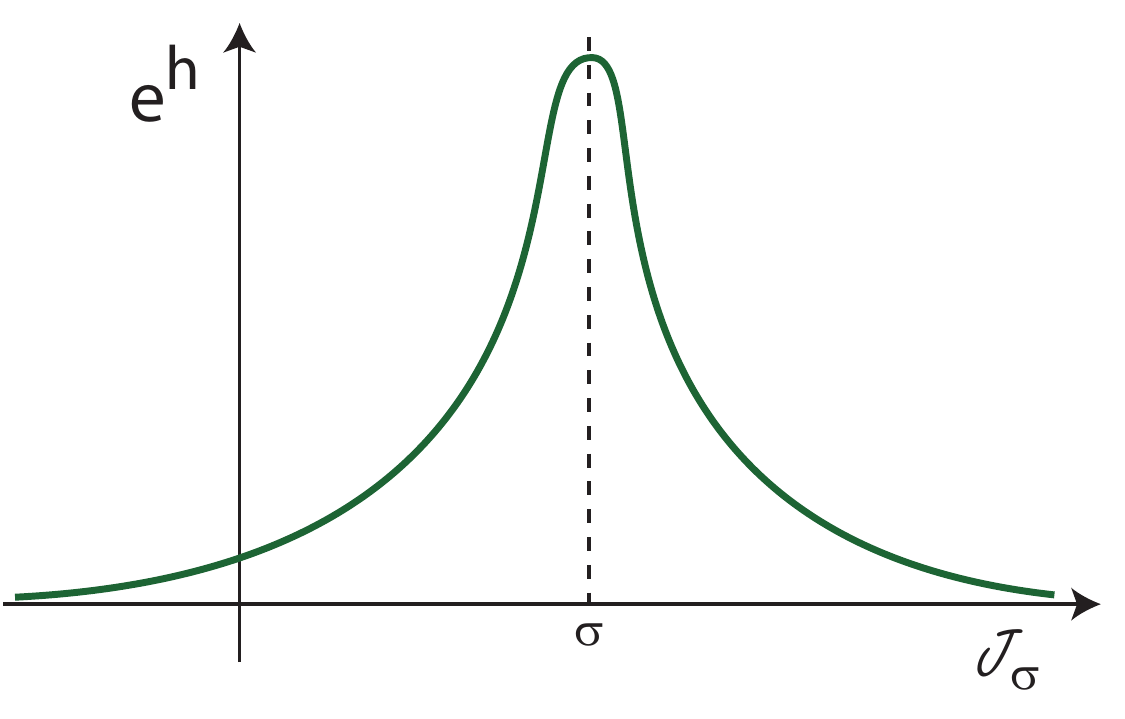}
	\end{minipage}%
	\caption{{\it Left panel:} From a saddle point $\sigma$ emanate upward (${\cal K}_\sigma$) and downward (${\cal J}_\sigma$) flows, which are located in the wedges $J_\sigma$ (in green) and $K_\sigma$ (in red) respectively, defined as the regions where the Morse function $h$ is lower (higher) than its value at the saddle, respectively. The arrows along the flows indicate the direction of descent, and the downward flow ${\cal J}_\sigma$ is known as a Lefschetz thimble. The wedges are separated by blue lines along which $h$ is constant and equal to the value at the saddle point $h(p_\sigma.)$  {\it Right panel:} Along a Lefschetz thimble the real part $h$ of the exponent decreases as fast as possible, ensuring an absolutely convergent integral.}
	\protect
	\label{fig:thimble}
\end{figure}

For the simple examples we discuss here, the obvious metric $\mathrm{d}s^2 = |\mathrm{d}x|^2$ is sufficient. Defining complex coordinates, $(u,\bar{u})=\bigl(($Re$([x]+i$Im$[x]),($Re$[x]-i$Im$[x])\bigr)$, the metric is $g_{uu}=g_{\bar{u}\bar{u}}=0,\,g_{u\bar{u}}=g_{\bar{u}u}=1/2$. Then $h=(\mathcal{I}+\bar{\mathcal{I}})/2$ and (\ref{eq:dw}) becomes 
\begin{equation}
\frac{\mathrm{d}u}{\mathrm{d}\lambda} = - \frac{\partial {\bar{\cal I}}}{\partial \bar{u}}, \quad \frac{\mathrm{d}\bar{u}}{\mathrm{d}\lambda} = - \frac{\partial {{\cal I}}}{\partial {u}}\,.
\end{equation} 
The imaginary part of the exponent $H = \text{Im}[iS/\hbar]$ is conserved along these flows, since 
\begin{equation}
\label{eq:imh}
\frac{\mathrm{d} H}{\mathrm{d}\lambda} = \frac{1}{2i}\frac{\mathrm{d}({\cal I} - \bar{\cal I})}{\mathrm{d}\lambda} = \frac{1}{2i}\left( \frac{\partial {\cal I}}{\partial u}\frac{\mathrm{d}u}{\mathrm{d}\lambda} - \frac{\partial \bar{\cal I}}{\partial \bar{u}}\frac{\mathrm{d}\bar{u}}{\mathrm{d}\lambda}\right) = 0\,.
\end{equation}
Thus the integrand $e^{iS[x]/\hbar}$ -- which was a purely oscillatory factor in the original integral -- does not oscillate at all when evaluated along a downward flow (see Fig.~\ref{fig:thimble}). Instead, it decreases monotonically so that the integral converges absolutely and ``as rapidly as possible.'' For a downward flow originating at a saddle, $\lambda$ runs from $-\infty$ at the saddle point to positive values as $h$ decreases. The Lefschetz thimble associated with a given saddle is defined as the set of downward flows leaving the saddle in this way. 

Analogously, upward flows are defined via
\begin{equation}
\frac{\mathrm{d}u^i}{\mathrm{d}\lambda} = +g^{ij}\frac{\partial h}{\partial u^j}\,,
\label{eq:uw}
\end{equation}
with $H$ likewise being conserved along these flows. Every critical point has an upward flow which, in analogy to the downward flow, is labelled ${\cal K}_\sigma$. 

There is a complication in this argument which it is convenient to resolve. It is possible, in principle, for a steepest descent contour from one saddle point $p_\sigma$ to terminate on another saddle point $p_\sigma'$, as $\lambda\rightarrow \infty$ so that it coincides with a steepest ascent contour from $p_\sigma'$.  Such a situation is generically unlikely, but it may occur as the result of a symmetry. For example, if $S[x]$ is a real function of $x$ (meaning it is real for all real $x$), then any complex saddle point necessarily comes with a complex conjugate partner. The Morse function $h = i(S[x]-S[\bar{x}])/(2\hbar)$ is generally different at these two saddles, whereas the imaginary part of the exponent $H$ is the same so that, indeed, both the steepest descent flow from the higher saddle and the steepest ascent flow from the lower saddle follow the line Re$[x]=const.$  Such a situation indeed arises with no boundary initial conditions, with the relevant contours exhibited in Fig.~5 below. 

Such a degeneracy between steepest ascent and steepest descent contours may generally be removed by adding an infinitesimal perturbation to $S[x]$, and defining the contour ${\cal C}$ in the limit as the perturbation is taken to zero. In this limit the contribution of the perturbation to the integral is negligible. However, a generic perturbation will break the degeneracy between the values of the imaginary part of the exponent $H$ at the two saddle points, making it impossible, according to (\ref{eq:imh}), for steepest ascent and descent flows from the two critical points to coincide. Of course, if a symmetry was responsible for the degeneracy, as we have discussed for the case where $S[x]$ is real, then the perturbation must violate the symmetry if it is to remove the degeneracy. So if $S[x]$ being real is responsible for the degeneracy, an imaginary perturbation will be needed to remove it. This is not a problem, however, since as explained, in the limit that the perturbation is taken to zero,  its influence on the integral is negligible.  

Once all such degeneracies are removed, we are left with a one-to-one correspondence between saddle points $p_\sigma$ and the associated steepest ascent and descent contours ${\cal J}_\sigma$ and ${\cal K}_\sigma$. The generic situation is then that any steepest descent contour from a saddle ends on a singularity where $h\rightarrow -\infty$, and any steepest ascent contour likewise ends on a singularity where $h\rightarrow +\infty$.

Thus, Lefshetz thimbles and upward flows only intersect at a single critical point, the one where both are defined. With a suitable choice of orientation, we can write for the intersection number
\begin{equation}
{\rm Int}({\cal J}_\sigma, {\cal K}_{\sigma'})=\delta_{\sigma \sigma'}. \label{eq:intersection}
\end{equation}
Our objective is to deform the original integral (\ref{eq:integral}) into one evaluated over a sum of Lefschetz thimbles. That is, we would like to write
\begin{equation}
\label{eq:contourexp}
{\cal C} = \sum_\sigma n_\sigma {\cal J}_\sigma,
\end{equation}
in a homological sense, for some integers $n_\sigma$ which may take the values $0$ or $\pm1$ when accounting for the orientation of the contour over each thimble. It follows from these equations that $n_\sigma= {\rm Int}(\mathcal{C}, {\cal K}_{\sigma})={\rm Int}(D, {\cal K}_{\sigma})$, since the intersection number is topological and will not change if we deform the contour ${\cal C}$ back to the original, real domain $D$. Thus a necessary and sufficient condition for a  given thimble ${\cal J}_\sigma$ to be relevant is that a steepest {\it ascent} contour from the critical point $p_\sigma$ intersects the original, real integration domain $D$. In this circumstance, intuitively, there is no obstacle to smoothly ``sliding'' the intersection point from the real axis along  ${\cal K}_{\sigma}$ down to $p_\sigma$, and in the process deforming the original integration contour onto the the thimble ${\cal J}_\sigma$. This is the argument we alluded to in the introduction, showing that if one starts from a real Lorentzian theory, one {\it never} obtains semiclassical enhancement factors such as are found in the Euclidean approach. 

In one complex dimension, the way this works is that the original integral along the real $x$-axis is deformed into a series of thimbles. With the appropriate choice of orientation, adjacent thimbles end and start on singularities of the Morse function $h$, so that there is no obstacle to deforming the combined contour back onto the real $x$-axis. The two ``free ends'' in the sum over thimbles, corresponding to the first and last steepest descent contours, from the first and last critical points, run to singularities of $h$ in a complex direction determined by the steepest descent flow. In order to show that the original integral $I$ equals the sum of integrals over thimbles, we must show that the original integration contour, which approaches the initial and final singularities of $h$ along the real $x$-axis, can be deformed into one which ends on initial and final steepest descent contours which approach the same singularities from a different direction. This requires that the integral taken along an ``arc'' drawn around the singularity vanishes in the limit that the arc is taken closer and closer to the singularity. 

We shall now illustrate this behavior in the integral which arises in the simplest models of minisuperspace quantum cosmology. As we shall see in the next section, this takes the form
\begin{equation}
\label{eq:exint}
\int_{0^+}^\infty {dN\over \sqrt{N} }e^{i f(N)/ \hbar}\,,
\end{equation}
where $f(N)$ is holomorphic in $N$ over the relevant domain.  The integrand possesses singularities at $N=0$ and $N=\infty$ and the contour of integration runs from one to the other, over all positive values of $N$. We wish to show that it is possible to deform this contour to a sum of the relevant steepest descent contours, the first and last of which approach the singularities of the integrand at some finite angle with respect to the real $N$-axis. 

Consider first a singularity of $f(N)$ which occurs at  infinite $N$.  We take the original integral up to some large positive value, $N_0$. It is convenient to change variables to $N=({\rm ln}z )^2$, so that (\ref{eq:exint}) becomes 2 $\int_1^{z_0} {dz\over z}e^{i f\left( (\ln z)^2\right)/ \hbar}$, with $z_0=e^{\sqrt{N_0}}$. The relevant steepest descent trajectory at large $|z|$ will be determined by the term with the largest power of $N$ in $f(N)$. It will run to infinity at some angle $\theta$ with respect to the real $z$-axis. To show that the original integral taken up to some large real value $z_0$ is accurately approximated by the steepest descent integral taken out to $(|z|,\theta)=(z_0, \theta_0)$, we need to show that the integral along an arc at fixed $|z|$ with the angle $\theta$ running from $0$ to $\theta_0$, becomes negligible as $z_0$ is taken to infinity. Assume, for example, that $f(N) = a N$ at large $|N|$, with $a$ positive. Now set $z=e^{\sqrt{N_0}+i \theta}$ so that the integral along the arc at fixed $|z|$ becomes
\begin{eqnarray}
\label{eq:exinta}
2 \int_0^{\theta_0} {dz\over z}e^{i a (\ln z)^2/ \hbar}=2 i \int_0^{\theta_0} d\theta e^{i a (\sqrt{N_0} +i \, \theta)^2/ \hbar}\equiv i I_0
\rightarrow |I_0|<2 \int_0^{\theta_0} d\theta e^{- 2 a \sqrt{N_0}\theta/ \hbar}<{\hbar\over a \, \sqrt{N_0}}, \label{eq:arcatinfinity}
\end{eqnarray}
where we used a standard Schwarz-type inequality, and the fact that the last integral is bounded by its value when taken over an infinite range.
We have thus bounded the magnitude of the integral along the arc at fixed $|z|$, by a quantity which tends to zero as $N_0$ tends to infinity. Hence in the limit of large $N_0$, the original contour may indeed be deformed to one ending on the steepest descent contour at the same value of $N_0$, with negligible change in the value of the integral. The limit $N_0\rightarrow \infty$ can now be taken, with the conclusion that the two integrals are identical in this limit. It is not hard to generalize this argument to any holomorphic $f(N)$ behaving as a power of $N$ at large $N$: one just needs to choose $N_0$ large enough to ensure that all terms in the real part of the exponent in the analog of (\ref{eq:exinta}) are bounded by some finite multiple of the term involving the highest power of $N_0$. 

Similarly, the steepest descent contour approaches the singularity at $N=0$ along a complex direction.  For example, if $f(N)\sim -a/N$  as $N\rightarrow 0$, with $a$ positive, then $N=0$ is approached from positive imaginary values. To show that the original integral  (\ref{eq:exinta}) taken along the real $N$-axis equals the steepest descent integral, we cut the former off at some small real $N=\epsilon_0$. Setting $N=1/(\ln z)^2$,  (\ref{eq:exinta}) becomes 2 $\int_1^{z_0} {dz\over z}{1\over (\ln z)^2}e^{- i a \left( (\ln z)^2\right)/ \hbar}$ with $z_0=e^{1/\sqrt{\epsilon_0}}$. By Cauchy's theorem, the original integral taken over $N>\epsilon_0$ may be deformed into an integral along an arc $z=e^{1/\sqrt{\epsilon_0}}e^{-i \theta}$, plus the steepest descent integral taken from the arc's intersection with the steepest descent contour. On the arc, $|\ln z|^2>1/\epsilon_0$, so the integral along the arc is bounded by $2 \epsilon_0  \int d\theta e^{2 a \theta/(\hbar \sqrt{\epsilon_0})}<\hbar \epsilon_0^{3\over 2}/a$ and hence vanishes as $\epsilon_0\rightarrow 0$. Therefore the steepest descent integral and the original Lorentzian integral give the same result in the limit as the cutoff is removed. 

Once we have deformed the contour from the real axis to run through a set of thimbles associated with the contributing critical points, we have:
\begin{equation}
\label{eq:contour}
I= \int_{D} \mathrm{d} x \, e^{iS[x]/\hbar} =\int_{\cal C} \mathrm{d} x \, e^{iS[x]/\hbar} = \sum_\sigma n_\sigma \int_{{\cal J}_\sigma} \mathrm{d} x \, e^{iS[x]/\hbar}.
\end{equation}
 As (\ref{eq:contour}) indicates, typically more than one Lefschetz thimble contributes to the Lorentzian path integral, with given boundary conditions, even in mini-superspace quantum cosmology. 

The integral taken over a thimble is absolutely convergent if
\begin{align}
\left| \int_{{\cal J}_\sigma} \mathrm{d} x e^{iS[x]/\hbar} \right| 
\leq \int_{{\cal J}_\sigma} |\mathrm{d}x| \left| e^{iS[x]/\hbar} \right| 
=
\int_ {{\cal J}_\sigma} \mathrm|{d}x| e^{h(x)} 
< \infty\,.
\label{eq:conv}
\end{align}
Defining the length along the curve as $l= \int |dx|$, the integral will converge if $h(x(l))< -\ln (l) + A$, for some constant $A$, as $l \rightarrow \infty$, which is a rather weak requirement. 

We have then expressed the original integral as a sum of absolutely convergent steepest descent integrals. In an expansion in $\hbar$, we have
\begin{equation}
I = \int_D \mathrm{d}x \, e^{i S[x]/\hbar} 
= \sum_\sigma n_\sigma \, e^{i \, H(p_\sigma)}\int_{{\cal J}_\sigma} e^h \mathrm{d}x 
\approx \sum_\sigma n_\sigma \, e^{i S(p_\sigma)/\hbar}\left[A_\sigma+\mathcal{O}(\hbar)\right],
\end{equation} 
where $A_\sigma$ represents the result of the leading-order Gaussian integral about the critical point $p_\sigma$. Sub-leading terms may be evaluated perturbatively in $\hbar$. In the case of degenerate $h$, a similar expansion applies -- we will encounter such an example later in the paper.

\section{Minisuperspace Lorentzian path integral} \label{section:HH}

In this paper we consider a universe with a positive cosmological constant $\Lambda$, described by the action
\begin{equation}
S = \frac{1}{2}\int_{\cal M} \mathrm{d}^4x \sqrt{-g} \left( R - 2 \Lambda\right) + \int_{\cal \partial M} \mathrm{d}^3y \sqrt{g^{(3)}} K\,,
\end{equation} 
where we have set $8\pi G = 1.$ The second term, involving the 3-metric $g^{(3)}_{ij}$ and the trace of the second fundamental form $K$ of the boundary $\partial {\cal M}$, is needed to ensure the variational principle yields the Einstein equations if the boundary geometries are held fixed. For simplicity, we truncate the theory to the simplest cosmologies, represented by the line element
\begin{equation}
\mathrm{d}s^2 = - N(t)^2 \mathrm{d}t^2 + a(t)^2 \mathrm{d}\Omega_3^2\,,
\label{eq:metric}
\end{equation}
with $\mathrm{d}\Omega_3^2$ the metric of a homogeneous, isotropic $3$-dimensional space with curvature $k$. This is a gross simplification of the original theory -- we no longer have propagating gravitational waves -- but we retain a dynamical scale factor $a(t)$ as well as diffeomorphism invariance in the timelike coordinate $t$, and these will be sufficient for us to illustrate many key features of Lorentzian quantum cosmology.   

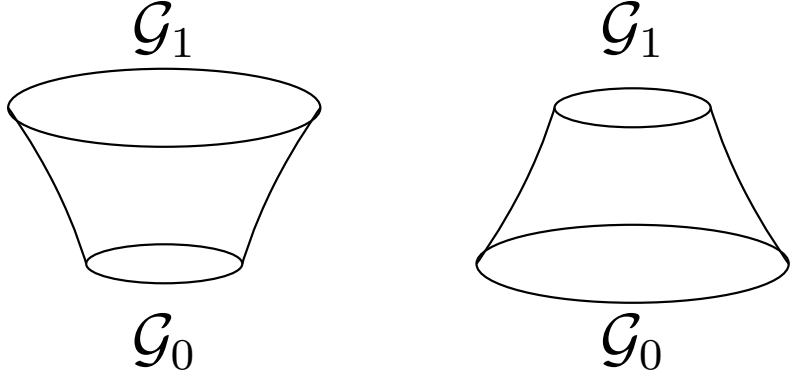
\begin{figure}
\centering
\resizebox {0.65\columnwidth}{!}{
\begin{tikzpicture}

\draw (0,0) circle [x radius=0.5, y radius=0.125];
\draw (0,1) circle [x radius=1, y radius=0.25];
\draw [black] plot [smooth, tension=1] coordinates { (-0.5,0) (-0.7,0.5) (-1,1)};
\draw [black] plot [smooth, tension=1] coordinates { (0.5,0) (0.7,0.5) (1,1)};

\draw (3,0) circle [x radius=1, y radius=0.25];
\draw (3,1) circle [x radius=0.5, y radius=0.125];
\draw [black] plot [smooth, tension=1] coordinates { (2.5,1) (2.3, 0.5) (2,0)};
\draw [black] plot [smooth, tension=1] coordinates { (3.5,1) (3.7,0.5) (4,0)};

\draw[] (0,-0.5) node[] {$\mathcal{G}_0$};
\draw[] (0,1.5) node[] {$\mathcal{G}_1$};

\draw[] (3,-0.5) node[] {$\mathcal{G}_0$};
\draw[] (3,1.5) node[] {$\mathcal{G}_1$};

\end{tikzpicture}}
\caption{A pictorial description of the Feynman propagator, with $\mathcal{G}_0$ and $\mathcal{G}_1$ the initial and final three-geometry. \textit{Left:} an expanding phase. \textit{Right:} a contracting phase.}
\label{fig:FeynmanProp}
\end{figure}

The Feynman path integral for the reduced theory is 
\begin{align}
G[a_1;a_0] = \int \mathcal{D} N \mathcal{D} \pi \mathcal{D} a \mathcal{D} p \mathcal{D} C \mathcal{D} \bar{P}\, e^{\frac{i}{\hbar} \int_0^1\left[ \dot{N}\pi + \dot{a}p + \dot{C}\bar{P}- N H\right]\mathrm{d}t}\,,
\end{align}
where, in addition to $a$, $N$ and the  fermionic ghost $C$, we have introduced the conjugate momenta $p,\pi$ and $\bar{P}$, and the corresponding Liouville measure. Without loss of generality, we can choose the range of the time coordinate to be $0\leq t\leq 1$. The Hamiltonian constraint $H[a,p;N,\pi;C,\bar{P}] = H_{EH}[a,p] + H_g[N,\pi;C,\bar{P}]$ consists of the Einstein-Hilbert Hamiltonian $H_{EH}$, in our case a minisuperspace Hamiltonian, and a Batalin, Fradkin and Vilkovisky (BFV) ghost Hamiltonian $H_g$\footnote{The Batalin, Fradkin and Vilkovisky ghost is an extension of the Fadeev-Popov ghost \cite{Batalin:1977}. The Fadeev-Popov ghost is based on the BRST symmetry. In particular, the constraint algebra forms a Lie algebra. In general relativity the constraint algebra does not close, which is why the BFV quantization is required. For minisuperspace we have only one constraint, $H$, for which the constraint algebra trivially closes. Thus the distinction is inessential here, but the BFV quantization is nevertheless preferable.}. The ghost is necessary since the minisuperspace action is diffeomorphism invariant. The ghost term breaks time reparametrization symmetry and fixes the proper-time gauge $\dot{N}=0$. For a detailed discussion of the BFV ghost in this setting see Teitelboim \cite{Teitelboim:1982,Teitelboim:1983} and Halliwell \cite{Halliwell:1988wc}. For minisuperspace models, most of the path integrals can be performed analytically, yielding
\begin{align}
G[a_1;a_0] = \int_{0^+}^\infty \mathrm{d}N \int_{a=a_0}^{a=a_1} \mathcal{D} a \, e^{i S(N,a)/\hbar}  \,,
\label{eq:gprop}
\end{align} 
which has a very simple interpretation. 
The path integral $\int \mathcal{D}  a e^{iS(N,a)/\hbar}$ represents the quantum mechanical amplitude for the universe to evolve from $a_0$ to $a_1$ in a proper time $N$. The integral over the lapse function indicates that we should consider paths of every proper duration $0< N<\infty$. Teitelboim \cite{Teitelboim:1983b} showed that this choice of integration domain leads to the causal ordering of the $a_0$ and $a_1$, {\it i.e.} $a_0$ precedes $a_1$. This allows us to describe both an expanding $a_1>a_0$ and a contracting $a_1< a_0$ universe, since the direction of the arrow of time is determined by the Feynman propagator and not by the choice of boundary conditions. For an illustration see Fig.~\ref{fig:FeynmanProp}.

The path integral $\int \mathcal{D}  a e^{iS(N,a)/\hbar}$ represents the amplitude for the universe to evolve from $a_0$ to $a_1$ in a proper time $N$. The integral over the lapse function indicates that we should consider paths $a(t)$ from $a_0$ to $a_1$, of every proper duration $0< N<\infty$. Teitelboim \cite{Teitelboim:1983b} showed that this choice of integration domain leads to the causal ordering of the $a_0$ and $a_1$, {\it i.e.} $a_0$ precedes $a_1$. This allows us to describe both an expanding $a_1>a_0$ and a contracting $a_1< a_0$ universe, since the direction of the arrow of time is determined by the Feynman propagator and not by the choice of boundary conditions. For an illustration see Fig.~\ref{fig:FeynmanProp}.

The action in (\ref{eq:gprop}) reduces to
\begin{equation}
S= 2 \pi ^2 \int_0^1 \mathrm{d}t N \left( - 3 a \frac{ \dot{a} ^2 }{N ^2} + 3ka - a ^3 \Lambda \right) \,. \label{eq:Action}
\end{equation} 
We are faced with a functional integral over $a(t)$, and an ordinary integral over the proper time $N$. The former may be performed in the semiclassical approximation. Notice first that the classical equations for $a(t)$  and $N$ are real. In fact, the equations of motion for $a(t)$ yield a unique solution in the form $a=a_c(Nt)$, for arbitrary initial and final $a_0$ and $a_1$. However, the constraints which follow from varying $N$ in generally can only be satisfied by complex $N$. Since  the equations of motion are real, it follows that solutions for $N$ come in complex  conjugate pairs. Also, reversing the sign of $N$ is classically equivalent to reversing the arrow of time, a symmetry of the classical equations. Hence we may anticipate that, quite generally, there  will be four solutions for $N$, with only two of them being distinct after time reversal symmetry is taken into account.  

In fact, we can simplify the calculation by noticing that redefining the lapse function $N(t)\rightarrow N(t)/a(t)$ renders the action (\ref{eq:Action})  
quadratic in $q(t) \equiv a(t)^2$, allowing the path integral over $q(t)$ to be performed exactly~\cite{Halliwell:1988wc}. (Actually, there is a subtlety since such a redefinition alters the path integral measure. More fundamentally, one must ensure that the quantum mechanical propagator is properly covariant under such changes of variable. As discussed in \cite{Halliwell:1988wc,Gielen:2016fdb}, the starting point for constructing the propagator is a proper ordering of the Hamiltonian operator. This ordering is determined by covariance under changes of variables including the one just given. With this correction, involving the Ricci curvature on superspace, the quantum Hamiltonians of the theories expressed in terms of $a$ or $q=a^2$ are equivalent. In the case of the redefinition considered here, the correction term is only important at small $a$. For simplicity, as well as consistency with earlier treatments, we shall ignore it in the leading, semiclassical analysis we perform in this paper.)

In these new variables, the action  (\ref{eq:Action}) becomes
\begin{equation}
S= 2 \pi ^2 \int_0^1 \mathrm{d}t \left( -\frac{3}{4 N}\dot{q}^2 + N(3k -  \Lambda q) \right) \,. \label{ActionH}
\end{equation} 
The equation of motion and the constraint following from this action are 
\begin{eqnarray}
\ddot{q}  =  \frac{2\Lambda}{3}N^2; \quad \frac{3}{4 N^2} \dot{q}^2 +3k  = \Lambda q \;. \label{qconstraint} 
\end{eqnarray}
With boundary conditions $q(0)=q_0$ and $q(1)=q_1,$ the general solution to the first equation (before imposing the constraint) is
\begin{equation}
\label{eq:classicalsolution}
\bar{q}=\frac{\Lambda}{3}N^2 t^2 + \left(- \frac{\Lambda}{3}N^2+ q_1- q_0\right) t + q_0\,.
\end{equation}
Writing the full solution, which does satisfy the constraint as
\begin{equation}
q(t) = \bar{q}(t) + Q(t)\,,
\end{equation}
the path integral becomes
\begin{equation}
G[q_1;q_0] = \int_0^\infty \mathrm{d} N e^{2\pi^2 i S_0/\hbar} \int_{Q[0]=0}^{Q[1]=0} \mathcal{D} Q e^{2\pi^2 i S_2/\hbar}\,,
\end{equation}
with
\begin{eqnarray}
S_0 &=&  \int_0^1 \mathrm{d}t \left( -\frac{3}{4 N}\dot{\bar{q}}^2 + 3kN - N \Lambda \bar{q} \right) \,, \label{ActionH4}\quad 
S_2 =  -\frac{3}{4 N} \int_0^1 \mathrm{d}t\,  \dot{Q}^2 \,. \label{ActionH3}
\end{eqnarray} 
The path integral over $Q$ is Gaussian and can be evaluated exactly:
\begin{equation} \label{Qintegral}
\int_{Q[0]=0}^{Q[1]=0} \mathcal{D} Q e^{2\pi^2 i S_2/\hbar} = \sqrt{\frac{3\pi i}{2N\hbar}} \,.
\end{equation}
The propagator thus reduces to an ordinary integral
\begin{equation}
\label{eq:propagator}
G[q_1;q_0] = \sqrt{\frac{3\pi i}{2\hbar}}\int_0^\infty \frac{\mathrm{d} N}{N^{1/2}} e^{2\pi^2 i S_0/\hbar}.
\end{equation}

Equation \eqref{eq:propagator} is an oscillatory integral, to which we apply the methods of the previous section. We lift the lapse $N$ to the complex plane and regard the boundary values $0$ and $\infty$ of the integral as points on the Riemann sphere. The action $S_0$ can be explicitly evaluated,
\begin{equation}
S_0 = N^3 \, \frac{\Lambda^2}{36} + N \left( -\frac{\Lambda}{2}(q_0+q_1) +3k \right) +\frac{1}{N}\left( -\frac{3}{4} (q_1-q_0)^2\right)\,.
\end{equation}

The action $S_0$ has four saddle points in the complex plane, which are solutions of
\begin{equation}
\partial S_0/\partial N= \Lambda^2 N_s^4  + \left( -6\Lambda (q_0+q_1) +36k \right) N_s^2  +9 (q_1-q_0)^2 = 0\,,
\end{equation}
given by
\begin{equation}
\label{Saddlesk1}
N_s = c_1 \frac{3}{\Lambda} \left[ \left( \frac{\Lambda}{3} q_0-k\right)^{1/2} + c_2 \left( \frac{\Lambda}{3} q_1-k\right)^{1/2}\right]\,,
\end{equation}
with $c_1,c_2\in\{-1,1\}$. The action evaluated at these saddle points is given by
\begin{eqnarray}
S_0^{saddle} &=& N_s^3 \, \frac{\Lambda^2}{36} + N_s \left( -\frac{\Lambda}{2}(q_0+q_1) +3k \right) +\frac{1}{N_s}\left( -\frac{3}{4} (q_1-q_0)^2\right) \nonumber\\
 &=& \frac{1}{N_s} \left[ N_s^4 \, \frac{\Lambda^2}{36} + N_s^2 \left( -\frac{\Lambda}{2}(q_0+q_1) +3k \right) - \frac{3}{4} (q_1-q_0)^2\right] \nonumber\\ 
 &=& \frac{1}{N_s} \left[  -\frac{\Lambda^2}{18}N_s^4 - \frac{3}{2}(q_1-q_0)^2 \right] \nonumber\\
&=& -c_1 \frac{6}{\Lambda}  \left[ \left(\frac{\Lambda}{3} q_0 - k\right)^{3/2} +c_2 \left(\frac{\Lambda}{3} q_1 - k\right)^{3/2} \right]\,,
\end{eqnarray}

Each of these four saddle points corresponds to a Lefschetz thimble $\{\cal J_\sigma\}$, and a steepest ascent contour $\{\cal K_\sigma\}$. Each is also associated with wedges $J_\sigma, K_\sigma$ in which the real part of the exponent $i S/\hbar$ is respectively lower and higher than the saddle point value. Writing the original integration contour in terms of the Lefschetz thimbles
\begin{align}
(0^+,\infty) = \sum_\sigma n_\sigma \cal J_\sigma\,,
\end{align}
we approximate the propagator using the saddle point approximation in the limit $\hbar \to 0$,
\begin{eqnarray}
G[q_1;q_0]  & = & \sum_\sigma n_\sigma \sqrt{\frac{3\pi i}{2\hbar}}\int_{{\cal J}_\sigma} \frac{\mathrm{d}N}{N^{1/2}} e^{2\pi^2 i S_0/\hbar} \nonumber\\ 
& \approx& \sum_\sigma n_\sigma\sqrt{\frac{3\pi i}{2\hbar}} \frac{e^{2\pi^2 i S_0^{saddle}/\hbar}}{N_s^{1/2}}\int_{{\cal J}_\sigma} \mathrm{d}N e^{\frac{i\pi^2}{\hbar}S_{0,NN}(N-N_s)^2} \left[ 1+ {\cal O}\left(\hbar^{1/2}\right)\right] \nonumber\\ 
& \approx& \sum_\sigma n_\sigma\sqrt{\frac{3\pi i}{2\hbar}} \frac{e^{2\pi^2 i S_0^{saddle}/\hbar}}{N_s^{1/2}}e^{i\theta_\sigma} \int_{{\cal J}_\sigma} \mathrm{d}n e^{-\frac{\pi^2}{\hbar}|S_{0,NN}|n^2} \left[ 1+ {\cal O}\left(\hbar^{1/2}\right)\right] \nonumber\\ 
& \approx& \sum_\sigma n_\sigma\sqrt{\frac{3 i}{2 N_s |S_{0,NN}|}} e^{i\theta_\sigma} e^{2\pi^2 i S_0^{saddle}/\hbar} \left[ 1+ {\cal O}\left(\hbar^{1/2}\right)\right]\,, \label{eq:spa}
\end{eqnarray}
where we defined $N-N_s \equiv ne^{i\theta}$ with $n$ real and $\theta$ being the angle of the Lefschetz thimble with respect to the positive real $N$ axis.

The intersection coefficient $n_\sigma$, the angle $\theta_\sigma$ and action at the saddle point $S_0^{saddle}$ all depend on the boundary conditions $q_0$ and $q_1$ and the spatial curvature $k$. In particular, saddle points can become relevant or irrelevant as the boundary conditions are varied. Earlier approaches  amount to choosing particular contour in the complex $N$ plane ``by hand,'' on the basis of some preconceived notions. However, as we argued in section \ref{sec:PL}, the virtue of the Lorentzian path integral combined with Picard-Lefschetz theory is that the proper combination of saddle points and relative phases between them is completely fixed.

As can be seen from equation \eqref{Saddlesk1}, for spherical three-geometries, $k=1$, the saddle points can be complex, while for the flat and hyperbolic case the saddle points are real. Complex saddle points imply non-classical behaviour since the propagator becomes dominated by non-Lorentzian geometries. In the following sections we concentrate on spherical expanding universes. We study the saddle point approximation \eqref{eq:spa} in four qualitatively different configurations:
\begin{itemize}
\item For $q_1 \geq q_0 > \frac{3}{\Lambda}$ the saddle points are real. These boundary conditions represent a classical universe. This case is studied in section \ref{section:classicalboundaryconditions}. 
\item For $q_1 > \frac{3}{\Lambda} > q_0$ one of the roots becomes imaginary. This case includes the ``no-boundary'' proposal and is studied in section \ref{section:noboundary}. 
\item The limiting case between the classical and quantum phase, for which $q_1\geq q_0= \frac{3}{\lambda}$, is studied in section \ref{sec:boundaryCase}.
\item  For $\frac{3}{\Lambda} > q_1 \geq q_0$ both square roots become imaginary and both $q_0$ and $q_1$ are expected to behave quantum mechanically. We study this case in section \ref{sec:nonClassical}.
\end{itemize}

\subsection{Classical boundary conditions} \label{section:classicalboundaryconditions}

\begin{figure} 
	\centering
	\includegraphics[width=0.75\textwidth]{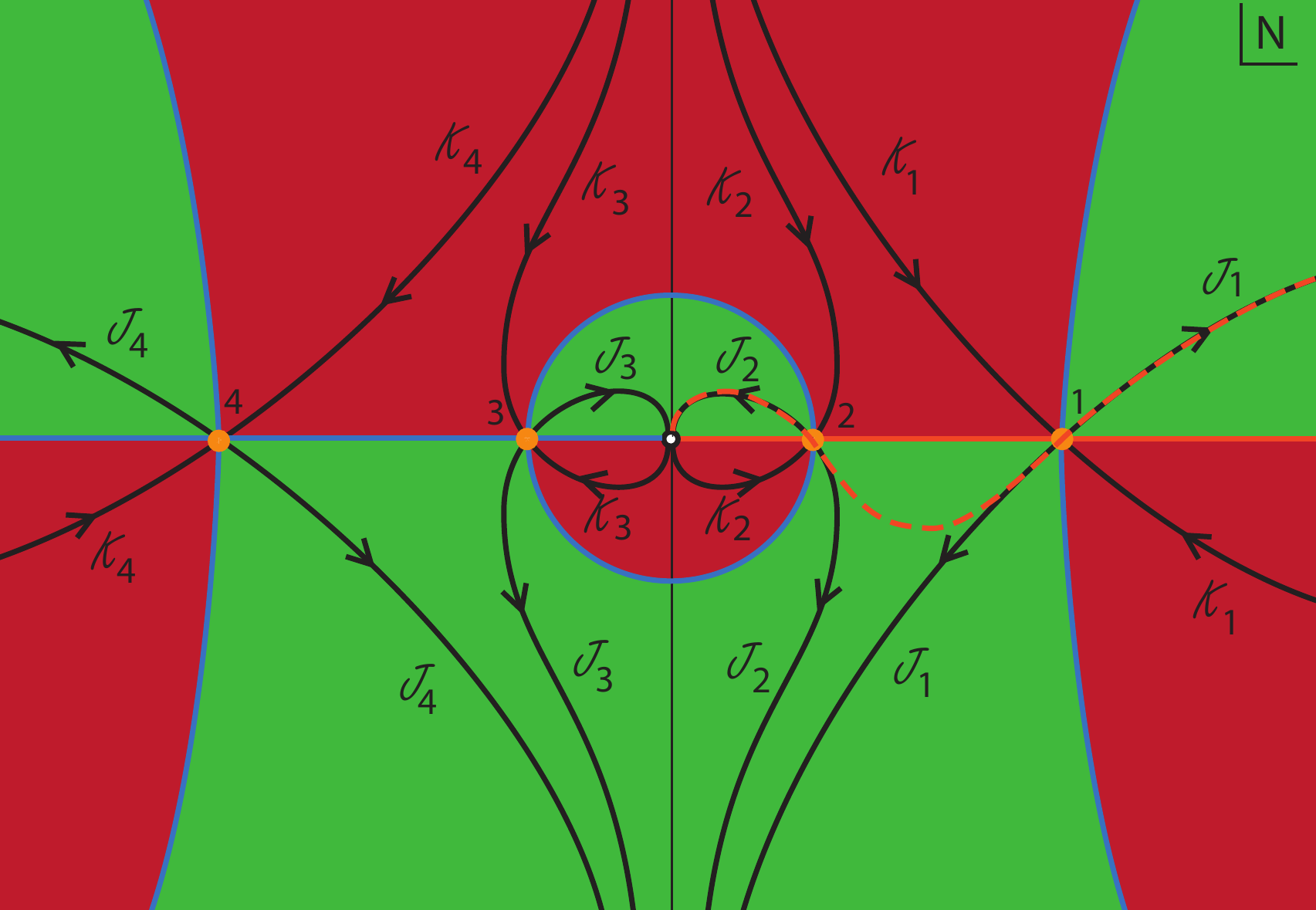}
	\caption{A sketch of the wedges and flow lines emanating from the saddle points in the complex $N$ plane, for classical boundary conditions $q_1>q_0>\frac{3}{\Lambda}$. The Lefschetz thimbles ${\cal J}_\sigma$ reside within the green wedges $J_\sigma$ (within which the magnitude of the integrand is smaller than at the corresponding saddle point), while the contours of steepest ascent ${\cal K}_\sigma$ reside within the red wedges $K_\sigma$ (within which the magnitude of the integrand is larger than at the corresponding saddle point). The arrows indicate the direction of steepest descent. The original integration contour along the positive real axis is shown in orange, and runs through two saddle points in this case. The deformed contour along which the integral is absolutely convergent comprises the thimbles ${\cal J}_1$ and ${\cal J}_2$: the dashed orange line indicates how the original contour is deformed onto to these thimbles. Note that neither the flow lines, nor the original integration contour, include the point at $N=0.$ }
	\protect
	\label{fig:classical2}
\end{figure} 

For classical boundary conditions $q_1 \geq q_0 > \frac{3}{\Lambda}$, the four saddle points are real, see Fig. \ref{fig:classical2} for the corresponding lines of steepest descent and ascent. The two positive saddle points 
\begin{equation}
N_{s\pm} = \sqrt{\frac{3}{\Lambda}} \left[ \left( q_1 - \frac{3}{\Lambda} \right)^{1/2} \pm \left( q_0 - \frac{3}{\Lambda} \right)^{1/2} \right]
\end{equation}
contribute to the integral since their curves of steepest ascent trivially intersects the original interval $(0,\infty)$. The two negative saddle points do not contribute to the propagator.

The equation of motion is solved by the de Sitter space solution. For classical boundary conditions there exist two solutions: either $q_0,q_1$ both sit on the same side of the waist of the de Sitter hyperboloid, or they are separated by the waist. Using the classical solution $\bar{q}(t)$ (see equation \eqref{eq:classicalsolution}) we can study the two saddle points. At $t=0$,
\begin{equation}
\frac{\mathrm{d}\bar{q}}{\mathrm{d}t} = -\frac{\Lambda}{3}N_{s\pm}^2 + q_1 - q_0 = 2\left(- q_0 + \frac{3}{\Lambda} \right) \mp 2 \sqrt{\left(q_0 - \frac{3}{\Lambda} \right) \left(q_1 - \frac{3}{\Lambda} \right) }\,.
\end{equation}
We observe that $N_{s+}$ corresponds to a decreasing solution. The waist of the de Sitter space sits between the specified boundaries. The other saddle point $N_{s-}$ corresponds to an increasing solution. In this case both boundaries sit on the same side of the waist. 

Figure \ref{fig:classical2} illustrates the Lefschetz thimbles corresponding to the saddle points. The first Lefschetz thimble runs from the origin at $N=0$ up in the positive imaginary $N$ direction\footnote{Note that this Lefschetz thimble does not include the point $N=0$ itself. In the small $N$ limit, the Morse function is approximated by $Re(-i/N)$ and this becomes arbitrarily negative as $N$ tends towards $N=0$ along the positive imaginary axis, without actually reaching $N=0.$ This is just as well, as the original integration contour also does not include $N=0$ at which point the metric would be singular.}, curves around, moves through $N_{s-}$, and asymptotically approaching the negative imaginary axis.  The second thimble runs up from the negative imaginary axis, through $N_{s+}$ and asymptotes to positive $Re(N)$ at an angle of $\pi/6$ \footnote{This angle is easy to determine: the flow lines correspond to lines of constant imaginary part of the integrand, and for large $N$ this means constant $Re(N^3)\,$.}. Note that the sum of these two thimbles is indeed deformable to the positive real $N$ axis. In Fig. \ref{fig:classical2} the integration contour that runs through the saddle points along the Lefschetz thimbles ${\cal J}_{1,2}$ is shown by a dashed orange line -- along this contour the integral is manifestly convergent. 

Since we have two relevant saddle points, the saddle point approximation of the propagator \eqref{eq:spa} is the sum of two phases,
\begin{eqnarray} \label{nbwf_classical}
G[q_1;q_0]  
&\approx& \left( \frac{3i}{4\Lambda \sqrt{(q_0-\frac{3}{\Lambda})(q_1-\frac{3}{\Lambda})}}\right)^{1/2}
 \left[ \, e^{-i\frac{\pi}{4}} e^{iS(N_{s-})/\hbar} +   \, e^{i\frac{\pi}{4}}e^{iS(N_{s+})/\hbar} \right] \nonumber\\
  &\approx& \frac{e^{i\frac{\pi}{4}}3^{1/2}}{[(\Lambda q_0 - 3)(\Lambda q_1 - 3)]^{1/4}}\cos\left(\frac{4\pi^2\Lambda^{1/2}}{3^{1/2} \hbar} \left(q_0 - \frac{3}{\Lambda}\right)^{3/2} -\frac{\pi}{4}\right) e^{-i \frac{4\pi^2\Lambda^{1/2}}{3^{1/2}\hbar}  \left(q_1 - \frac{3}{\Lambda}\right)^{3/2}}\,, 
\end{eqnarray} 
The factors $e^{\pm i\frac{\pi}{4}},$ arise from aligning the fluctuation integrals with the Lefschetz thimbles (cf. the orange dashed line), as explained in more detail in section~\ref{section:noboundary}. 

In this simple model, we have interference effects between the two possible classical trajectories linking our initial and final conditions. In more realistic models involving interactions with other fields, one might expect the two terms in the transition amplitude to decohere. This would suppress the interference.

\subsection{No-boundary conditions} \label{section:noboundary}

\subsubsection{Implications of Picard-Lefschetz theory}

The ``no-boundary'' conditions were proposed by Hartle and Hawking as a theory of initial conditions for the universe \cite{Hawking:1981gb,Hartle:1983ai,Hawking:1983hj}. The idea is that in the path integral one should sum only metrics whose only boundary is provided by the final spatial hypersurface (corresponding to the current state of the universe). To implement ``no-boundary'' conditions, we must take $q_0=0$ and find a 4-metric which is regular there. This is possible for positive $k$. The ``no-boundary'' condition is supplemented with the constraint equation \eqref{qconstraint} evaluated at $q=0,$ 
\begin{equation}
\dot{q}^2 = -4N^2 k \qquad (q=0)\,\,.
\end{equation}
We will take the final boundary to correspond to a late time configuration, where the universe has become large, $q_1 > \frac{3}{\Lambda}.$ The saddle points of the action are given by
\begin{equation} \label{Saddlesnb1}
N_{s,nb1}=+ \frac{3}{\Lambda} \left[ i \pm \left(\frac{\Lambda}{3} q_1 - 1\right)^{1/2} \right]\,, \quad N_{s,nb2}=- \frac{3}{\Lambda} \left[ i \pm \left(\frac{\Lambda}{3} q_1 - 1\right)^{1/2} \right] \,,
\end{equation}
with corresponding actions
\begin{equation}
S_{0,nb1} = - \frac{6}{\Lambda}  \left[-i \pm \left(\frac{\Lambda}{3} q_1 - 1\right)^{3/2} \right]\,, \quad S_{0,nb2} = + \frac{6}{\Lambda}  \left[-i \pm \left(\frac{\Lambda}{3} q_1 - 1\right)^{3/2} \right].
\end{equation}
Note that saddle points in the upper half plane lead to a $e^{i 2\pi^2 S_0} \sim e^{-12\pi^2/(\hbar\Lambda)},$ while those in the lower half plane lead to $e^{i 2\pi^2 S_0} \sim e^{+12\pi^2/(\hbar\Lambda)}.$
\begin{figure} 
	\centering
	\includegraphics[width=0.6\textwidth]{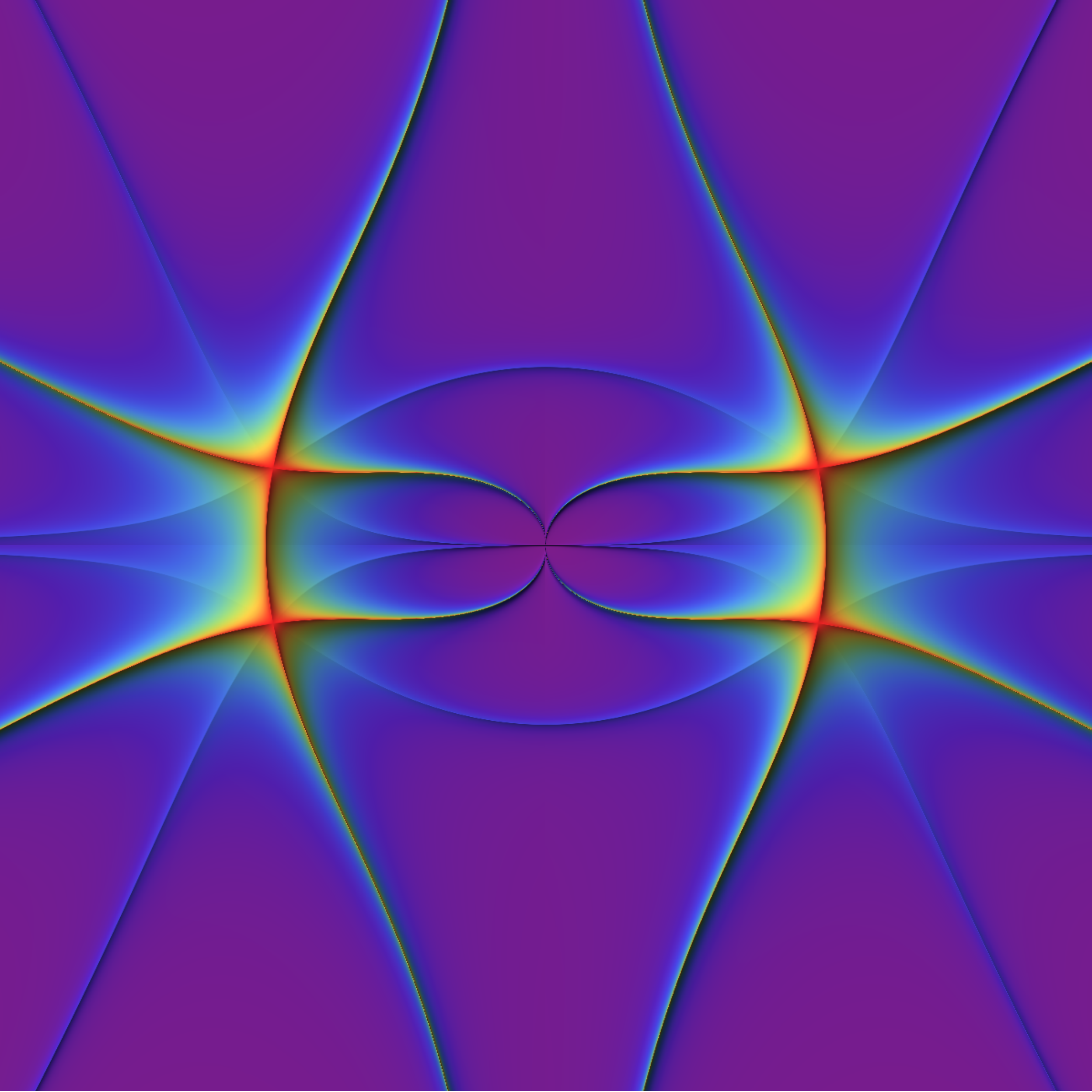}
	\caption{For this numerical example we have chosen $k=1, \Lambda = 3, q_0=0, q_1=10.$ The saddle points then lie at $\pm 3 \pm i.$ Shown in the present figure are both the boundaries of wedges (lines of constant real part of the integrand/imaginary part of the action -- light blue lines) and the flow lines (lines of constant real part of the action -- red/green lines). More specifically, the plot shows both $Abs[{\rm Im}(S(N) - S(N_s))]$ and $Abs[{\rm Re}(S(N) - S(N_s))]$, where lighter colours correspond to smaller values. The four saddle points are located at the intersections of the flow lines. More details are provided in Fig. \ref{fig:upper}.}
	\protect
	\label{fig:saddleS}
\end{figure} 

\begin{figure} 
	\centering
	\includegraphics[width=0.75\textwidth]{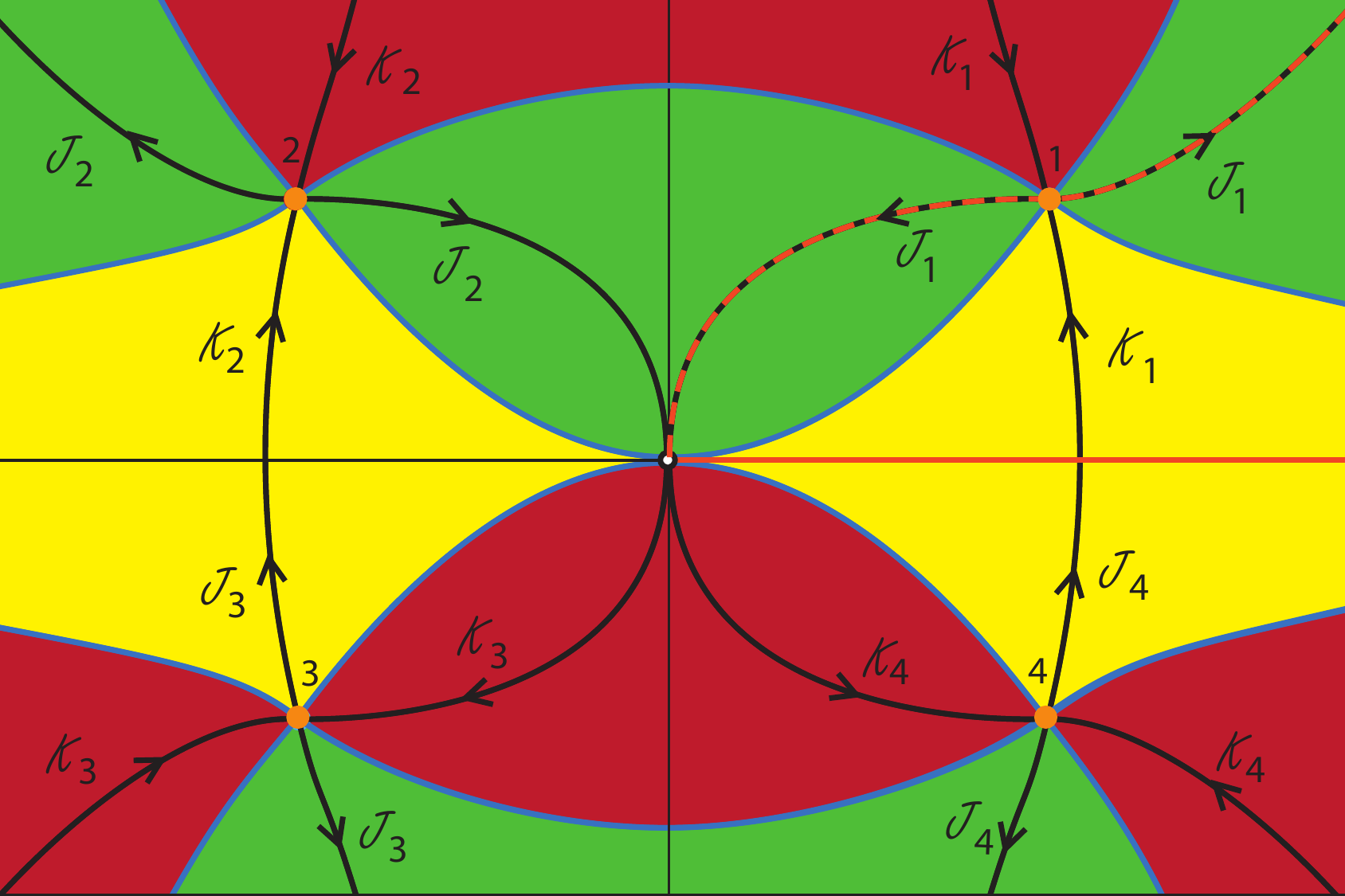}
	\caption{A sketch of the wedges and flow lines emanating from the saddle points in the complex $N$ plane, for ``no-boundary'' conditions $q_0=0, q_1 > \frac{3}{\Lambda}.$ The loci of the steepest ascent/descent flows (in black) and of the boundaries between wedges (in blue) were determined numerically in Fig. \ref{fig:saddleS}. Here the arrows indicate the direction of steepest descent. We have coloured the wedges such that regions $J_\sigma$ with a lower value of the magnitude of the integrand than the corresponding saddle point are green, and regions $K_\sigma$ with a higher value are red, with the exception of the yellow regions which have a value intermediate between the two saddle point values. Comparing with the adjacent colours then avoids any ambiguity. Notice that, due to the symmetry explained above equation (\ref{eq:intersection}), there are `degenerate' ascent and descent flows that link saddle points. This degeneracy is broken by adding an infinitesimal perturbation to the action, as shown in Fig.~\ref{fig:perturbation}. The original integration contour along the positive real axis is shown in orange, and the deformed contour which Picard-Lefschetz theory picks out as the preferred integration cycle is marked in dashed orange. Again neither the flow lines, nor the original or final integration contours, include the point at $N=0.$ Only saddle point $1$ in the upper right quadrant can be linked to the original integration contour via an upward flow, and this implies that the (orange-dashed) downward flow from this saddle point is the correct Lefschetz thimble along which the path integral should be performed. }
	\protect
	\label{fig:upper}
\end{figure}

Given the saddle points, we can determine the wedges and the curves of steepest descent and ascent emanating from them. We use the fact that curves with Re$(iS_0)$ specify the boundaries of the wedges, and that Im$(iS_0)$ is constant along the flow lines to determine them numerically -- see also \cite{Halliwell:1988ik}. For the case of interest to us, the wedge boundaries and flow lines are shown in Fig. \ref{fig:saddleS}, while the directions of the flows are sketched in Fig. \ref{fig:upper}. 

\begin{figure} 
	\centering
	\includegraphics[width=0.6\textwidth]{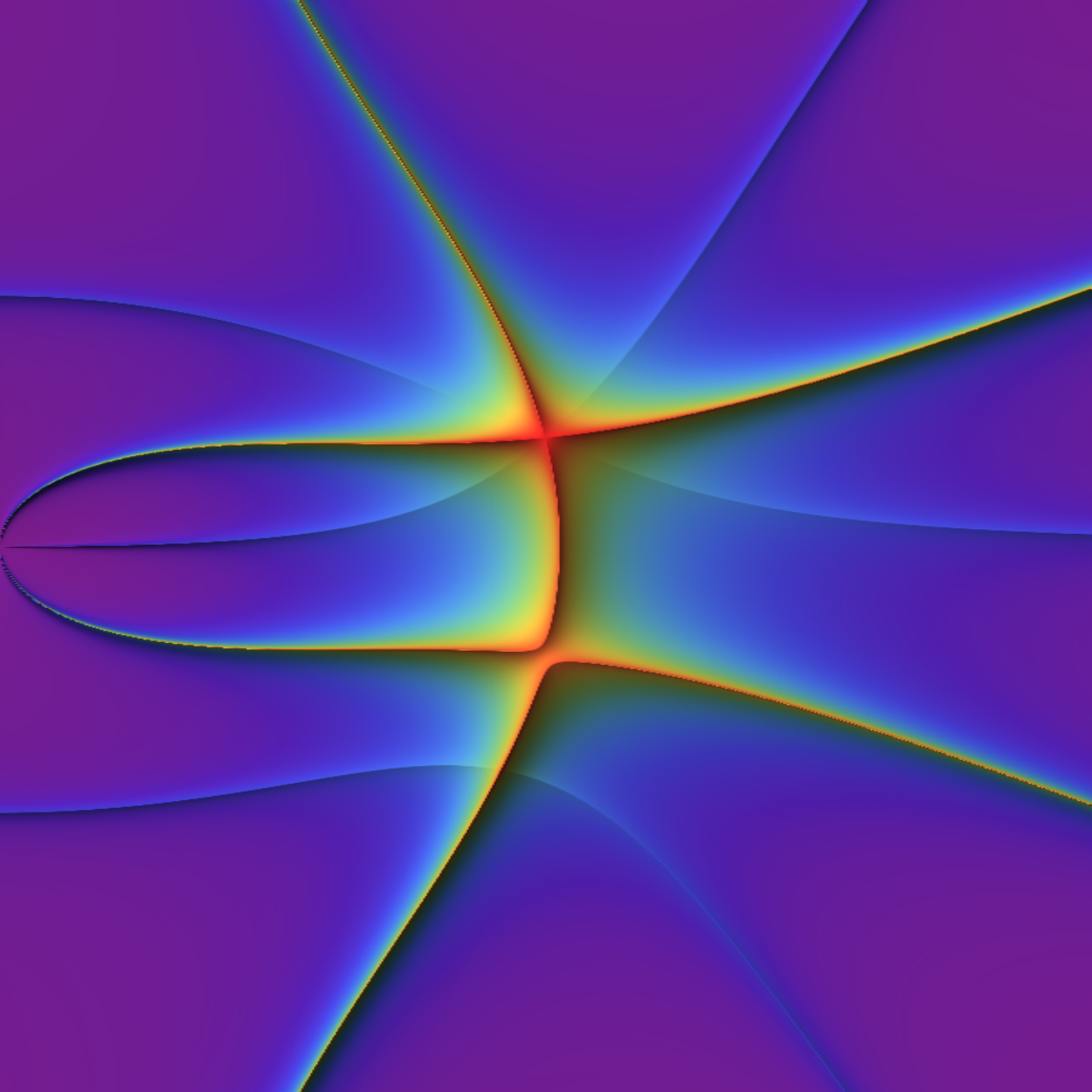}
	\caption{Due to the reality of the action, discussed above equation (\ref{eq:intersection}), and as shown in Fig. \ref{fig:upper}, a curve of steepest ascent from saddle point $1$ coincides with a curve of steepest descent from saddle point $4.$ This degeneracy can be lifted by adding a small complex perturbation to the action, in this example $\Delta S_0 = i N/100.$  The new contours of steepest ascent and descent, as well as the level sets of the magnitude of the integrand are shown for saddle point $1$. In the presence of the the perturbation, the lower steepest ascent contour from saddle point $1$, instead of joining saddle point $4$, now runs left to join the origin $N=0$ from below. Hence the formula for the intersection number \eqref{eq:intersection} may now be used to unambiguously determine that saddle point $1$ is relevant to the Lorentzian path integral. Note that the two lower curves, one green/red and one blue, do not now pass through a saddle point and should be ignored. }
	\protect
	\label{fig:perturbation}
\end{figure}

One can identify the direction of the flows analytically by expanding the action around a saddle point,
\begin{equation}
\delta S_0 = S_0^{saddle} + \frac{1}{2} S_{0,NN}^{saddle} (\delta N)^2 + \dots
\end{equation}
The second derivative is given by
\begin{eqnarray}
S_{0,NN}^{saddle} &=& N_s \, \frac{\Lambda^2}{6} - \frac{3}{2N_s^3}(q_1-q_0)^2 \\ &=&  \frac{1}{6 N_s^3} \left[\Lambda^2 N_s^4 - 9 (q_1-q_0)^2 \right]\,.
\end{eqnarray}
By evaluating the second derivative, we can determine the direction along which the imaginary part of the action stays constant. Our numerical example serves as an illustration. For the saddle point (number $2$) at $N_s=-3+i$ for instance, the second derivative is given by $S_{,NN}=9/5 \times(-1 + 3i).$ This means that $\alpha \equiv Arg(S_{,NN}) = \pi - ArcTan(3) \approx 1.89.$ We want $Im(i S - i S(N_s))=0.$ Around the saddle point, the change in $iS$ goes like $\Delta(iS) \propto i S_{,NN} (\delta N)^2 \sim n^2 e^{i(\pi/2 + 2\theta +\alpha)},$ where we have written $\delta N = n e^{i\theta}.$ The change in the imaginary part will be proportional to $\sin(\pi/2 + 2\theta +\alpha),$ and this change is zero if 
\begin{eqnarray} \label{angle}
\theta &=& \frac{k\pi}{2} - \frac{\pi}{4} - \frac{\alpha}{2}, \qquad k \in \mathbb{R} \\
&\approx& -0.16,\, 1.41,\, 2.98,\, 4.55,\, \dots
\end{eqnarray} 
This is in good agreement with the flow lines shown in the figure. Note that the change in the real part of the integrand $h=Re(iS)$ is given by $\cos(\pi/2 + 2\theta +\alpha).$ The direction of steepest descent is thus given by $\cos(\pi/2 + 2\theta +\alpha)=-1,$ i.e. for 
\begin{eqnarray} \label{descentangle}
\theta &=&  k\pi + \frac{\pi}{4} - \frac{\alpha}{2}, \quad k \in \mathbb{R}, \quad (steepest \,\, descent) \\
&\approx& -0.16,\, 2.98,\, \dots
\end{eqnarray}
while the curves of steepest ascent are at $\theta \approx 1.41,\, 4.55,\, \dots$
Thus the line of steepest descent of $h$ is towards the origin $N=0,$ while the curve of steepest ascent is down towards the real line. This line eventually connects with the saddle point at $-3-i$. 
Thus we encounter the degenerate situation described in section \ref{sec:PL} where the curves of steepest ascent from one saddle point coincides with the curve of steepest descent from another. As discussed there, we can lift this degeneracy by considering a small complex perturbation of the action, and subsequently take the limit where the perturbation vanishes. The effect of such a perturbation is shown in Fig. \ref{fig:perturbation}, where one can clearly see that the degeneracy is now lifted, and the intersection formula \eqref{eq:intersection} can be applied. 
It is straightforward to repeat this calculation for the other three saddle points, with the result that the two saddles in the upper half plane have flow lines that are mirror images of each other, while the two saddles in the lower half plane have their upward and downward flow reversed.  

Note that the downward flow lines (Lefschetz thimbles) of the upper saddle points can indeed be deformed to the real $N$ line, while the downward flow lines of the lower saddle points cannot. Moreover, only saddle point $1$ can be linked to the original integration contour (the positive real half line) via an upward flow, and hence the appropriate integration contour, along which the integral will be manifestly convergent, is given by the Lefschetz thimble ${\cal J}_1$ also indicated by the dashed orange line in Fig. \ref{fig:upper}. 
More precisely, it is implied by the arguments presented around Eq. \eqref{eq:arcatinfinity} that the integral along the arc at infinity linking the real integration domain to the Lefschetz thimble ${\cal J}_1$ vanishes, and thus the path integral manifestly converges. 
Saddle point $1$ lies at
\begin{equation} 
N_{s,nb1}^+=+ \frac{3}{\Lambda} \left[ i + (\frac{\Lambda}{3} q_1 - 1)^{1/2} \right] \,,
\end{equation}
and the action evaluated on the saddle point is
\begin{equation}
S_{0,nb1}^+ = - \frac{6}{\Lambda}  \left[-i + (\frac{\Lambda}{3} q_1 - 1)^{3/2} \right]\,.
\end{equation}
For saddle points of the form \eqref{Saddlesk1}, we have 
\begin{equation}
S_{0,NN} = \frac{2c_2}{N_s} \left( \Lambda q_0 -3 \right)^{1/2} \left( \Lambda q_1 -3 \right)^{1/2}\,,
\end{equation}
implying that $Arg(N_s) = -\alpha + Arg\left[ \left( \Lambda q_0 -3 \right)^{1/2} \left( \Lambda q_1 -3 \right)^{1/2}\right].$ For the ``no-boundary'' conditions we thus find $Arg(N_s) + \alpha = \frac{\pi}{2},$ and combined with \eqref{descentangle} this implies $\theta - \frac{1}{2}Arg(N_s)=0$.
In the saddle point approximation, we thus obtain the wavefunction 
\begin{equation} \label{nbwf}
G_{nb}[q_1;0]   \approx e^{i\frac{\pi}{4}}\frac{3^{1/4}}{2 (\Lambda q_1-3)^{1/4}}e^{-12\pi^2/(\hbar \Lambda) -i 4\pi^2 \sqrt{\frac{\Lambda}{3}} (q_1 - \frac{3}{\Lambda})^{3/2}/\hbar }\,.
\end{equation}
Note that the real part of the classical action for the dominant saddle point is negative, as expected from the general arguments presented in section \ref{sec:PL}. This concludes the explicit derivation of our result that the relevant saddle point contributes a weighting $e^{-12\pi^2/(\hbar \Lambda)}$, the inverse of the Hartle-Hawking result. 

\subsubsection{Classicality}

The properties of the physical spacetime should be inferred from the quantum mechanical amplitude. In particular, whether or not we are describing a classical spacetime depends on how the amplitude changes as its arguments are varied. Above we have calculated $G_{nb}[q_1]$ as a function of $q_1,$ the scale factor on a spatial hypersurface. The amplitude $G_{nb} = e^{A + i P}$ that we have obtained has a slowly varying amplitude $A$ and a fast-varying phase $P$ as the universe expands, i.e. in the large $q_1$ limit
\begin{equation}
\frac{\partial A/\partial q_1}{\partial P/\partial q_1} \sim \frac{1}{(q_1 - \frac{3}{\Lambda})^{3/2}} \rightarrow 0\,.
\end{equation} 
This implies that the amplitude is increasingly classical in a WKB sense as the universe expands. Hence it describes a classical universe. The scaling of the WKB condition for large $q_1$ is inversely proportion to the volume of space since the spatial volume is proportional to $q_1^{3/2}$. This is what is expected from studies of inflationary ``no-boundary'' instantons in the limit of an exactly flat potential \cite{Lehners:2015sia}. 

\subsubsection{Relation to the Euclidean path integral}

It is interesting to ask why our results differ from the earlier approaches that took as their starting point the Euclidean path integral. After all, one can simply translate our results into this language by replacing the lapse function $N$ by $iN$. The graphs we plotted would then simply rotate by $90$ degrees. Why would any physical results be changed? The crucial point is that the Euclidean approach assumes the Euclidean time to be fundamental. The path integral should really be performed along the imaginary $N$ axis. In other words, in the Euclidean approach one would take the original integration contour to extend from $N=0$ (again excluding the point at $N=0$ itself) to infinity in the positive or negative \emph{imaginary} direction. At this point it is useful to take another look at Fig. \ref{fig:upper}. First note that {\it none} of the saddle points are related to the imaginary axis by an upward flow line. There are two flow lines that tend towards $N=0$ asymptotically, but they do not intersect the imaginary axis. This immediately implies that one cannot perform the integral thus defined using the saddle point method. In other words, no combination of saddle points provides a good estimate of the value of the integral. What is more, the integral has no chance of converging. In the positive imaginary direction, the integral diverges at large values of $i|N|,$ while along the negative imaginary axis it diverges as it approaches $N=0.$ Another way to say this is to observe that every integration path containing a saddle point passes through a region where the integral is divergent, when one tries to smoothly deform it to the imaginary axis using Cauchy's theorem. Hence we conclude that the Euclidean path integral is simply ill-defined. By contrast, the real time path integral leads to unambiguous and convergent results.     

\subsection{Boundary conditions at the classical limit}
\label{sec:boundaryCase}

\begin{figure} 
	\centering
	\includegraphics[width=0.75\textwidth]{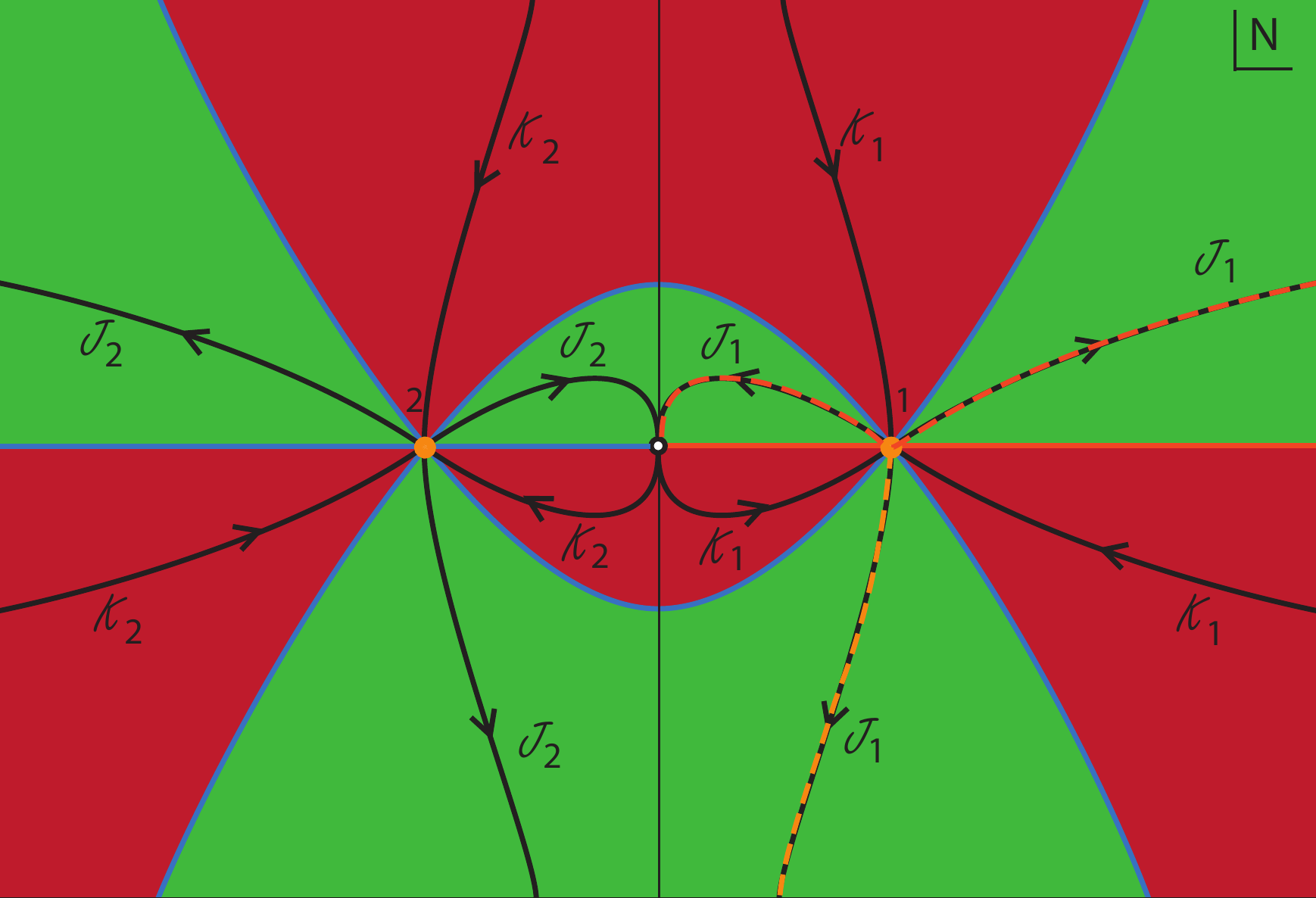}
	\caption{A sketch of the wedges and flow lines emanating from the saddle points in the complex $N$ plane, for the boundary conditions $q_1 > q_0=\frac{3}{\Lambda}$. The colours and arrows are as described in the caption of Fig. \ref{fig:classical2}. For these boundary conditions, the saddle points are degenerate, and there are three lines of steepest ascent and descent emanating from them.}
	\protect
	\label{fig:limit2}
\end{figure}

The saddle points and flow lines in the case of ``no-boundary'' conditions look rather different than those obtained with classical boundary conditions. One may wonder how the two descriptions link up as the boundary conditions are continuously varied from classical to non-classical, i.e. from $q_0> 3/\Lambda$ to $q_0 < 3/\Lambda,$ while keeping the final condition classical, $q_1> 3/\Lambda$. Here we consider limiting case $q_1 > q_0=\frac{3}{\Lambda}$. The wedges, flow lines and their description are given in Fig. \ref{fig:limit2}.

The action $iS_0$ has two degenerate saddle points at
\begin{equation}
N_{s,limit\pm} = \pm \sqrt{\frac{3}{\Lambda}} \left( q_1 - \frac{3}{\Lambda} \right) ^{1/2}\,.
\end{equation}
The saddle point $N_{s,limit+}$ lies on the original integration contour, and will contribute to the path integral. The other saddle point $N_{s,limit -}$ is irrelevant to the propagator. The saddle points are of order $2,$ since 
\begin{equation}
\frac{\partial S_0}{\partial N} = \frac{\partial^2 S_0}{\partial N^2} = 0 \quad \text{at} \quad N=N_{s,limit\pm}\,,
\end{equation}
while the third derivative is non-zero. This means that the Taylor expansion around the saddle point is dominated by a cubic term,
\begin{equation}
S_0 = S_0(N_{s,limit+}) + \frac{1}{6}\frac{\partial^3 S_0}{\partial N^3}(N_{s,limit+}) (\delta N)^3 + \dots
\end{equation}
This explains why the flow lines now intersect at angles of $\pi/3.$ It is straightforward to evaluate the third derivative,
\begin{eqnarray}
S_{0,NNN}^{saddle} &=& \frac{\Lambda^2}{6} + \frac{9}{2N_s^4}(q_1-q_0)^2 \\ &=& \frac{6}{N_s^4}\left[\left( q_0-\frac{3}{\Lambda}\right)^2 + \left( q_1-\frac{3}{\Lambda}\right)^2 \right]\,.
\end{eqnarray}
In our case $S_{0,NNN}(N_{s,limit+})=2\Lambda^2/3.$ Given that the third derivative is real-valued and positive, we can determine the directions of the upward and downward flow, as indicated in Fig. \ref{fig:limit2}. 

This helps us to understand the contour of integration. Starting from $N=0$, the contour follows the Lefschetz thimble ${\cal J}_1$ and moves in the positive imaginary direction, and passes through the saddle point with positive real part. It then runs down, asymptotically  towards negative imaginary values, and subsequently comes back along the same path, crossing the degenerate saddle point once more before shooting off at an angle of $\pi/6$ with respect to the real $N$ axis.  The middle part of the contour sums to nothing. The Lefschetz thimble transitions to being located entirely in the upper half plane when the boundary condition on $q_0$ becomes non-classical, cf. Fig. \ref{fig:upper}.

Since for the boundary condition $q_0=\frac{3}{\Lambda}$ the function $h$ is not a Morse function, i.e. $h$ has two degenerate critical points, the saddle point approximation of equation \eqref{eq:spa} does not apply. However, using the integral $\int_{0}^{+\infty} \mathrm{d}n e^{-k \, n^3}  = \Gamma(\frac{4}{3})k^{-1/3},$ we can approximate
\begin{eqnarray}
\int \frac{\mathrm{d}N}{N^{1/2}}e^{ikN^3} &\approx& \frac{1}{N_s^{1/2}} \int \mathrm{d}[\delta N]e^{ik(\delta N)^3} \nonumber \\ 
&\approx&  \frac{1}{N_s^{1/2}} \left[ e^{-i\frac{\pi}{6}} \int_{- \infty}^0 \mathrm{d}n e^{+kn^3} + e^{+i\frac{\pi}{6}} \int_0^\infty e^{-kn^3}\right] \nonumber\\ 
&\approx& \frac{\sqrt{3} \Gamma(\frac{4}{3})}{N_s^{1/2}k^{1/3}}\,, 
\end{eqnarray}
in order to obtain the saddle point approximation of the propagator
\begin{equation} \label{nbwf_classical_limit}
G[q_1;q_0=3/\Lambda]  \approx \frac{e^{i\frac{\pi}{4}}3^{17/12}\Gamma(\frac{4}{3})}{2^{5/6}\pi^{1/6}(\hbar\Lambda)^{1/6}(\Lambda q_1-3)^{1/4}} \, e^{-i4\pi^2\sqrt{\frac{\Lambda}{3}} (q_1-\frac{3}{\Lambda})^{3/2}/\hbar} \,. 
\end{equation} 
The prefactors $e^{\pm i \frac{\pi}{6}}$ arise from writing $\delta N = n e^{\pm i \frac{\pi}{6}}$ so that the respective integrals are performed along the Lefschetz thimbles.

\subsection{Non-classical boundary conditions}
\label{sec:nonClassical}

Finally, we consider boundary conditions that are classically impossible, $\frac{3}{\Lambda}>q_1 \geq q_0$, where both scale factors are smaller then the waist of de Sitter space. Even though such configurations are impossible in Lorentzian signature, they exist in Euclidean signature and correspond to sections of a 4-dimensional sphere (e.g. one may picture them as surfaces of constant latitude). Correspondingly, the saddle points are pure imaginary. In the upper half plane they are
\begin{equation}
N_{upper\pm} = i \sqrt{\frac{3}{\Lambda}} \left[ \left( \frac{3}{\Lambda} - q_1 \right)^{1/2} \pm \left(\frac{3}{\Lambda} -q_0 \right)^{1/2} \right]\,,
\end{equation}
while in the lower half plane
\begin{equation}
N_{lower\pm} = - i \sqrt{\frac{3}{\Lambda}} \left[ \left( \frac{3}{\Lambda} - q_1 \right)^{1/2} \pm \left(\frac{3}{\Lambda} -q_0 \right)^{1/2} \right]\,.
\end{equation}
There are two possibilities, as for the case of classical boundary conditions. The two spatial hypersurfaces are either on one side of the equator of the 4-sphere, or are separated by the equator. We can determine which is which by looking at the derivative of the classical solution
\begin{equation}
\frac{\mathrm{d}\bar{q}}{\mathrm{d}t} =\frac{2\Lambda}{3}N_s^2 t -\frac{\Lambda}{3}N_s^2 + q_1 - q_0 = 0, \quad 0 < t < 1\,.
\end{equation}
It is straightforward to see that for $N_{upper+}$ and $N_{lower+}$ the scale factor squared $q$ reaches a maximum for $0<t<1,$ while for $N_{upper-}$ and $N_{lower-}$ the maximum is only reached for $t>1,$ which is outside of the range for which the solution has been determined. Thus  the saddle points at $N_{upper+}$ and $N_{lower+}$ correspond to the configuration where the initial and final hypersurface lie on different sides of the equator of the sphere. 

\begin{figure}[h] 
	\includegraphics[width=0.75\textwidth]{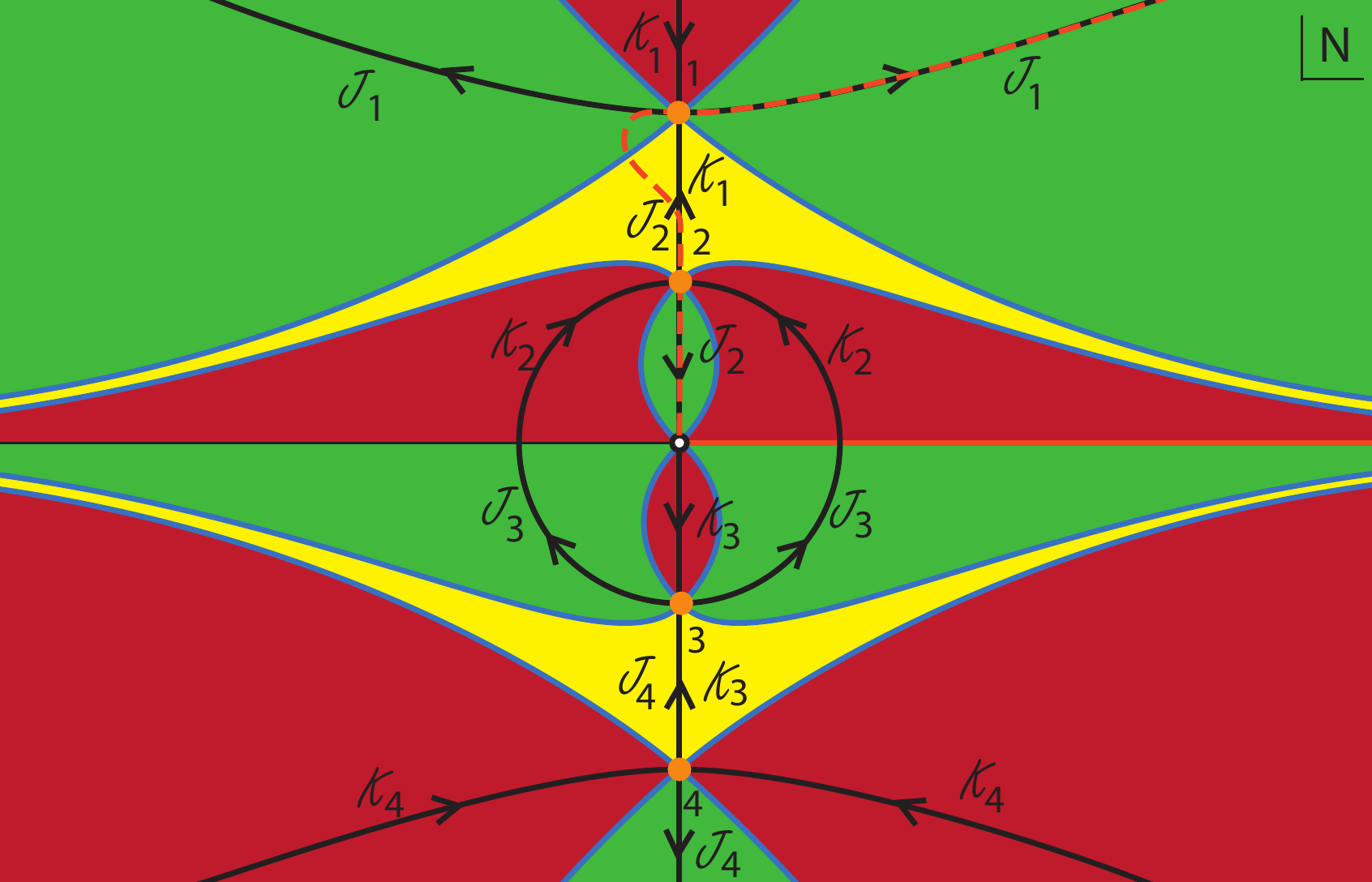}
	\caption{A sketch of the wedges and flow lines emanating from the saddle points in the complex $N$ plane, for non-classical boundary conditions $\frac{3}{\Lambda}>q_1 \geq q_0$. The lines, arrows and colours are as described in the caption of Fig. \ref{fig:upper}. Here we again have yellow wedges that are both higher than one saddle point and lower than another one, $J_2 = K_1$ and $J_4 = K_3.$ The original integration contour can be deformed into the dashed preferred contour by flowing down. Note that the preferred, dashed contour is chosen to follow the paths of steepest descent near the saddle points, so as to facilitate a saddle point approximation of the propagator.}
	\protect
	\label{fig:nonclassical1}
\end{figure}

An analysis of the Lefschetz thimbles indicates that only the saddle points in the upper half plane contribute to the propagator, as these are related to the original integration contour via an upward flow (in the case of $N_{upper+}$ the flow proceeds through the saddle point $N_{upper-}$ this degeneracy can once again be removed by adding a small complex perturbation to the action, and letting it vanish at the end of the calculation). The proper integration contour runs up from $N=0$ in the positive imaginary direction, passes through the two upper saddle points (along paths of steepest descent) and then follows the Lefschetz thimble in the upper right quadrant off to infinity at an asymptotic angle of $\pi/6$ with respect to the real $N$ axis -- see Fig. \ref{fig:nonclassical1}. The propagator can thus be approximated as
\begin{equation} \label{nbwf_nonclassical}
G[q_1;q_0]  \approx c_- e^{-\frac{12\pi^2}{\hbar\Lambda}\left[\left(1-\frac{\Lambda q_0}{3}\right)^{3/2}+\left(1-\frac{\Lambda q_1}{3}\right)^{3/2}\right]} +  c_+ e^{-\frac{12\pi^2}{\hbar \Lambda}\left[\left(1-\frac{\Lambda q_0}{3}\right)^{3/2}-\left(1-\frac{\Lambda q_1}{3}\right)^{3/2}\right]}  \,.
\end{equation}
Keeping in mind that for the saddle point $N_{upper-}$ we have the angle along the Lefschetz thimble being $\theta = \frac{\pi}{2}$ and thus $e^{i\left[\theta - \frac{1}{2}Arg(N)\right]}=e^{i\frac{\pi}{4}}$, while for $N_{upper+}$ we have $\theta = 0$ and $e^{i[\theta - \frac{1}{2}Arg(N)]}=e^{-i\frac{\pi}{4}}$, the normalization constants are
\begin{eqnarray}
c_\pm &=& e^{\mp i\frac{\pi}{4}} \left( \frac{3i}{4\Lambda \sqrt{\left(\frac{3}{\Lambda}-q_0\right)\left(\frac{3}{\Lambda}-q_1\right)}}\right)^{1/2}\,.
\end{eqnarray} 
Note that the exponents in the amplitude are purely real, so it most certainly does not describe classical cosmological evolution. The implied interference between the two terms in the wavefunction may lead to interesting effects, whose investigation we leave to future work. Here we simply note that the saddle point at $N_{upper-}$ dominates, {\it i.e.}, the path integral gives a higher weighting to the smaller 4-geometry connecting initial and final boundary values, and not the geometry for which the equator of the $4$-sphere sits in between. In particular, in the limit where the two boundaries become equal the preference is for having a vanishing $4$-geometry, a physically reasonable result.


\section{Relation to the Wheeler-DeWitt equation} \label{sectionWdW}

So far we have focussed on path integral quantization. It is natural to wonder how our results are related to canonical methods. In the canonical approach, one obtains a time independent Schr\"{o}dinger-type equation on superspace, {\it i.e.}, on the space of 3-dimensional spatial geometries. This is the Wheeler-DeWitt equation. Below we will derive this equation in two separate ways, first from the canonical approach, and afterwards by starting from the path integral. Both approaches lead to the same equation, as they must. The advantage of the path integral approach, however, is that it incorporates the correct boundary conditions and leaves no residual ambiguity in the solutions of the Wheeler-DeWitt equation. 

\subsection{Canonical derivation of the Wheeler-DeWitt equation}
The minisuperspace model is described by the action \eqref{ActionH}, with the corresponding Lagrangian
\begin{align}
L = 2\pi^2\left[ - \frac{3}{4N}\dot{q}^2 + 3kN - N \Lambda q\right].
\end{align}
The canonical momentum corresponding to $q$ is given by
\begin{align}
p = \frac{\partial L}{\partial \dot{q}} = -   \frac{3 \pi^2}{N}  \dot{q}\,.
\end{align}
The classical Hamiltonian of the system takes the simple form
\begin{align}
H =& \dot{q} p -L= -\frac{N}{6 \pi^2} \left[ p^2 + 12 \pi^4 (3k -\Lambda q)\right]=N\hat{H}\,.
\end{align}
The phase-space representation of the action reads
\begin{align}
S = \int \left(\dot{q}p - N\hat{H}\right)\mathrm{d}t = \int \left(\dot{q}p + \frac{N}{6 \pi^2} \left[ p^2 + 12 \pi^4 (3k -\Lambda q)\right]\right)\mathrm{d}t\,.
\end{align}
Observe that the lapse $N$ is a Lagrange multiplier leading to the classical constraint
\begin{align}
\hat{H}=0\,.
\end{align}

\indent In the canonical quantization scheme we obtain the Hamiltonian operator $\hat{H}$ in the $q$-representation by the substitution $p \mapsto \hat{p} = - i  \frac{\partial}{\partial q}$,
\begin{align}
\hat{H} =& \frac{1}{6\pi^2}\left[ \frac{\partial^2}{\partial q^2} + 12 \pi^4 (\Lambda q -3k)\right].
\end{align}
The corresponding Wheeler-DeWitt equation is given by
\begin{align}
 \hat{H} \psi = 0 \rightarrow  \,\, & \hbar^2\frac{\partial^2 \psi}{\partial q^2} + 12 \pi^4 (\Lambda q - 3k)\psi =0\,,
\end{align}
with $\psi$ the wave-function of the universe. The corresponding Feynman propagator $G$ satisfies \cite{Halliwell:1988wc}
\begin{align}
\hat{H}G = - i \delta(q_0-q_1)\,,
\end{align}
where the Hamiltonian operator acts either on $q_0$ or on $q_1$. We call this equation the inhomogeneous Wheeler-DeWitt equation. For a the De Sitter universe,
\begin{align}
 & \hbar^2\frac{\partial^2 G}{\partial q_1^2} + 12 \pi^4 (\Lambda q_1 - 3k)G = - 6 \pi^2 i \delta(q_0-q_1)\,.
\end{align}
In the next section we prove that the path integral is propagator satisfying this equation.

\subsection{Integral representation of the Feynman propagator}
Our discussion in section \ref{section:HH} has shown that the propagator/wave function of the universe is given by
\begin{align}
G[q_1;q_0] = \int_0^\infty \mathrm{d}N G[q_1;q_0;N]\,.
\end{align} 
The integrand
\begin{align}
G[q_1;q_0;N]= \int_{q=q_0}^{q=q_1}\mathcal{D}q e^{iS(N,q)/\hbar}
\end{align}
is the propagator to propagate from $q_0$ to $q_1$ in parameter time $N$. In analogy with non-relativistic quantum mechanics and by construction of the path integral, the propagator will satisfy a Schr\"odinger-like equation
\begin{align}
i\frac{\partial G[q_1;q_0;N]}{\partial N} = \hat{H} G[q_1;q_0;N]\,,
\end{align}
with the Hamiltonian operator acting on either $q_0$ or $q_1$ with the boundary condition
\begin{align}
\lim_{N\to 0} G[q_1;q_0;N] = \delta(q_0-q_1)\,.
\end{align}
It follows that the total propagator of minisuperspace models is a Greens function of the Wheeler-DeWitt equation
\begin{align}
\hat{H}G[q_1;q_0]
=&\int_0^\infty \mathrm{d}N \hat{H}G[q_1;q_0;N]\nonumber\\
=&i\int_0^\infty \mathrm{d}N \frac{\partial G[q_1;q_0;N]}{\partial N}\nonumber\\
=&i G[q_1;q_0;N]\big|_{N=0}^{N=\infty} \nonumber\\
=& - i \delta(q_0-q_1)\,.
\end{align}

\subsection{Solution of the Feynman propagator}
The Wheeler-DeWitt equation is solved by bilinear expressions in Airy functions, 
\begin{equation}
Ai\left[ \frac{\sqrt[3]{-12 \pi^4}(q \Lambda - 3k)}{\Lambda^{2/3}}\right], \quad Bi\left[ \frac{\sqrt[3]{-12 \pi^4}(q \Lambda - 3k)}{\Lambda^{2/3}}\right]\,,
\end{equation}
where $q$ stands for either $q_0$ or $q_1$ here. Note that in the arguments above, $\sqrt[3]{-1}$ can stand for $e^{i\frac{\pi}{3}}, -1, e^{-i\frac{\pi}{3}}$ -- either choice, combined with a suitable linear combination of Airy functions, solve the Wheeler-DeWitt equation \footnote{This is due to the fact that Airy functions contain three regions of convergence in the complex plane of the argument, where these regions are invariant under rotations by $2\pi/3$ radians. The Airy functions $Ai$ and $Bi$ are defined as two linearly independent combinations of the possible convergent contours. Rotating the arguments by $2\pi/3$ then translates into taking different linear combinations of the convergent integration contours, i.e. simply corresponds to taking different linear combinations of Airy functions \cite{Witten:2010cx}. An example is provided by the relation $Ai(z e^{\pm i \frac{2\pi}{3}}) = \frac{1}{2}e^{\pm i \frac{\pi}{3}}[Ai(z) \mp i Bi(z)]$.}. For definiteness we will choose $\sqrt[3]{-1}= e^{i\frac{\pi}{3}},$ and write
\begin{equation}
z \equiv  \frac{e^{i\frac{\pi}{3}}{(12 \pi^4)^{1/3}}(\Lambda q - 3)}{(\hbar\Lambda)^{2/3}}\,.
\end{equation}
We also use the notation $z_0 = z(q \rightarrow q_0)$ and $z_1 = z(q \rightarrow q_1)$.\\
\indent In the previous discussion, we approximated the Feynman propagator using the WKB approximation. Now, using the homogeneous solutions of the Wheeler-DeWitt equation we can construct the space of all possible Green's functions and subsequently determine the Feynman propagator. The propagators of the Wheeler-DeWitt equations must be of the form
\begin{align}
G[q_1;q_0] = - 6\pi^2 i \frac{\psi_1(q_0)\psi_2(q_1)\theta(q_0-q_1) + \psi_1(q_1)\psi_2(q_0)\theta(q_1-q_0) }{\psi_1(q_0)\psi_2'(q_1)-\psi_1(q_1)\psi_2'(q_0)},
\end{align}
with $\psi_1$ and $\psi_2$ two linearly independent homogeneous solutions. Explicitly, if we take $\psi_1 = Ai[z] + a Bi[z]$ and $\psi_2 = Ai[z] + b Bi[z]$ then we obtain the general form of the propagator
\begin{align}
G[q_1;q_0] = \frac{i}{(b-a)}\frac{2^{1/3}(-3)^{2/3} \pi^{5/3}}{(\Lambda\hbar)^{1/3}}\left[\psi_1(q_0)\psi_2(q_1) \theta(q_0-q_1)+ \psi_1(q_1)\psi_2(q_0)\theta(q_1-q_0)\right].
\end{align}

We can try to find the appropriate linear combination of solutions by solving for the path integral directly near $z_0=z_1=0,$ i.e. near $q_0 = q_1 = \frac{\Lambda}{3}$ \cite{Halliwell:1988ik}. Re-scaling the lapse function by $\Lambda^{2/3}$ and writing $z_0=z_1=Z,$ the path integral is given by
\begin{equation}
G[Z,Z] = \sqrt{\frac{3\pi i}{2}}\frac{1}{\Lambda^{1/3}} \int \frac{\mathrm{d}N}{N^{1/2}} e^{\frac{i \pi^2}{18}N^3 - \gamma NZ}
\end{equation}
where $\gamma = \frac{2\pi^2i}{(-12\pi^4)^{1/3}}.$ In order to find the Taylor expansion near $Z=0,$ we must evaluate the path integral and its first few derivatives at $Z=0.$ Writing $N=n e^{i\frac{\pi}{6}}$ in order to match to the appropriate Lefschetz thimble, we find
\begin{eqnarray}
G\left[\frac{\Lambda}{3},\frac{\Lambda}{3}\right] &=& \sqrt{\frac{3\pi i}{2}}\frac{1}{\Lambda^{1/3}} e^{i\frac{\pi}{12}} \int_0^\infty \frac{\mathrm{d}n}{n^{1/2}} e^{\frac{- \pi^2}{18}n^3} = \sqrt{\frac{3\pi i}{2}}\frac{1}{\Lambda^{1/3}}\frac{(2i)^{1/6}\Gamma(\frac{1}{6})}{(9\pi)^{1/3}} \\ 
\frac{\partial G}{\partial Z}\left[\frac{\Lambda}{3},\frac{\Lambda}{3}\right] &=& \sqrt{\frac{3\pi i}{2}}\frac{(-\gamma)}{\Lambda^{1/3}} e^{i\frac{\pi}{4}} \int_0^\infty \mathrm{d}n \, n^{1/2} e^{\frac{- \pi^2}{18}n^3} = \sqrt{\frac{3\pi i}{2}}\frac{(-\gamma)}{\Lambda^{1/3}} e^{i\frac{\pi}{4}} \sqrt{\frac{2}{\pi}} \\ 
\frac{\partial^2 G}{\partial Z^2}\left[\frac{\Lambda}{3},\frac{\Lambda}{3}\right] &=& \sqrt{\frac{3\pi i}{2}}\frac{\gamma^2}{\Lambda^{1/3}} e^{i\frac{5\pi}{12}} \int_0^\infty \mathrm{d}n \, n^{3/2} e^{\frac{- \pi^2}{18}n^3} =  \sqrt{\frac{3\pi i}{2}}\frac{\gamma^2}{\Lambda^{1/3}} e^{i\frac{5\pi}{12}} \frac{2^{5/6}3^{2/3}\Gamma(\frac{5}{6})}{\pi^{5/3}}
\end{eqnarray}
The Taylor series $G[Z,Z] = G\left[\frac{\Lambda}{3},\frac{\Lambda}{3}\right] + G_{,Z}\left[\frac{\Lambda}{3},\frac{\Lambda}{3}\right] Z + \frac{1}{2}G_{,ZZ}\left[\frac{\Lambda}{3},\frac{\Lambda}{3}\right] Z^2$ precisely matches that of 
\begin{equation}
G[Z,Z] =\frac{2^{1/3}(-3)^{2/3}\pi^{5/3}}{(\hbar\Lambda)^{1/3}} Ai[Z] \left( Ai[Z] - i Bi[Z] \right)\,.
\end{equation}
The asymptotic limits at large and small $q$, to which we will turn shortly, then imply that we should take
\begin{align}
G[q_1;q_0] =\frac{2^{1/3}(-3)^{2/3}\pi^{5/3}}{(\hbar\Lambda)^{1/3}} \bigg[&\left( Ai[z_0] - i Bi[z_0] \right)  Ai[z_1] \theta(q_1-q_0) \nonumber\\
&+\left( Ai[z_1] - i Bi[z_1] \right)  Ai[z_0] \theta(q_0-q_1)\bigg]\,.
\label{eq:prop}
\end{align}
Note that $a = -i$ and $b=0$, by which we see that \eqref{eq:prop} is a propagator. This is equal the evaluation of a non-relativistic particle with a linear potential \cite{Grosche:1998}. 

For the ``no-boundary'' case, we are interested in small (or zero) $q_0$ and large $q_1.$ For $\Lambda q_0 \ll 3$ we have the approximate formulae 
\begin{eqnarray}
Ai\left[ e^{i\frac{\pi}{3}}\frac{{(12 \pi^4)^{1/3}}}{\Lambda^{2/3}}(\Lambda q_0-3)\right]  \!\!&\approx&\!\!  \frac{(\hbar\Lambda)^{1/6}e^{-i\frac{\pi}{12}}}{\sqrt{\pi}(12\pi^4)^{1/12}(3-\Lambda q_0)^{1/4}} \sin\left(\! i \frac{4\pi^2}{\sqrt{3}\hbar\Lambda}(3-\Lambda q_0)^{\frac{3}{2}} \!+\! \frac{\pi}{4} \! \right) \\  
Bi\left[ e^{i\frac{\pi}{3}}\frac{{(12 \pi^4)^{1/3}}}{\Lambda^{2/3}}(\Lambda q_0-3)\right]  \!\!&\approx&\!\!  \frac{(\hbar\Lambda)^{1/6}e^{-i\frac{\pi}{12}}}{\sqrt{\pi}(12\pi^4)^{1/12}(3-\Lambda q_0)^{1/4}} \cos\left(\! i \frac{4\pi^2}{\sqrt{3}\hbar\Lambda}(3-\Lambda q_0)^{\frac{3}{2}} \!+\! \frac{\pi}{4} \! \right) 
\end{eqnarray}
while for $\Lambda q_1 \gg 3$ we have 
\begin{eqnarray} Ai\left[ e^{i\frac{\pi}{3}}\frac{{(12 \pi^4)^{1/3}}}{\Lambda^{2/3}}(\Lambda q_1-3)\right]  &\approx& \frac{(\hbar\Lambda)^{1/6}e^{-i\frac{\pi}{12}}}{2\pi^{1/2}(12\pi^4)^{1/12}(\Lambda q_1 - 3)^{1/4}} e^{-i\frac{\sqrt{12}\pi^2}{\hbar\Lambda}(\Lambda q_1 - 3)^{3/2}} \,.
\end{eqnarray}
For the total propagator we obtain
\begin{eqnarray} 
G[q_1;q_0=0] &=& \frac{2^{1/3}(-3)^{2/3}\pi^{5/3}}{(\hbar\Lambda)^{1/3}} \left(Ai\left[-3\frac{e^{i\frac{\pi}{3}}{(12 \pi^4)^{1/3}}}{(\hbar\Lambda)^{2/3}}\right] - i Bi\left[ -3\frac{e^{i\frac{\pi}{3}}{(12 \pi^4)^{1/3}}}{(\hbar\Lambda)^{2/3}}\right]\right) Ai[z_1] \nonumber \\
  &\approx& \frac{e^{i\frac{\pi}{4}}3^{1/4}}{2(\Lambda q_1 - 3)^{1/4}}  e^{-\frac{12\pi^2}{\hbar\Lambda} \, - \, i\frac{\sqrt{12}\pi^2}{\hbar\Lambda}(\Lambda q_1 - 3)^{3/2}} \,.
\end{eqnarray}
This agrees exactly with \eqref{nbwf}, including the subleading terms. 

A similar agreement can be found for classical and non-classical boundary conditions, by evaluating the appropriate limits of the Airy functions. In both cases the expressions agree with Eq. \eqref{nbwf_classical} respectively Eq. \eqref{nbwf_nonclassical}, including the first sub-leading term. 

With boundary conditions at the classical limit, $q_0 = \frac{3}{\Lambda}$ and $\Lambda q_1 \gg 3,$ the propagator is given by
\begin{align}
G\left[q_1;q_0=\frac{3}{\Lambda}\right] =\frac{2^{1/3}(-3)^{2/3}\pi^{5/3}}{(\hbar\Lambda)^{1/3}} \left( Ai[0] - i Bi[0] \right)  Ai[z_1] \,.
\end{align}
Using the exact expression $Ai[0] - i Bi[0]  = e^{i\frac{2\pi}{3}}\frac{2 \ 3^{1/3}}{\Gamma(-\frac{1}{3})} = e^{-i\frac{\pi}{3}}\, \frac{3^{5/6}\Gamma(\frac{4}{3})}{\pi},$ one may verify that the asymptotic form of this propagator also agrees with the path integral expression \eqref{nbwf_classical_limit}.

\section{Discussion} \label{sectionConclusions}

We hope the present paper has brought a new element of rigor into quantum cosmology. We have argued that the Lorentzian path integral, combined with Picard-Lefschetz theory, is to be preferred over the Euclidean version. In particular, in the simplest cosmology -- a closed FRW universe, with a positive cosmological constant -- we explained how it eliminates the ambiguities associated with the Euclidean path integral, including the conformal factor problem and the question of which saddle points are relevant. We have shown that Picard-Lefschetz theory identifies precisely which saddle point solutions contribute to the Feynman propagator, with which factors, and hence how a consistent semiclassical expansion may be developed. We have also shown how the path integral formulation of the causal propagator eliminates the problem of defining boundary conditions on superspace, an ambiguity which plagues attempts to obtain the ``wavefunction of the universe'' by solving the homogeneous Wheeler-DeWitt equation. 

As we have seen, for ``no-boundary'' conditions the Feynman propagator includes a semiclassical factor $e^{-12\pi^2/(\hbar \Lambda)}$, arising from the classical action of the relevant saddle point solution. We gave a general argument in the introduction, detailed in section (\ref{sec:PL}), that relevant complex classical solutions only give suppression factors, and never enhancement factors such as are obtained from Hartle and Hawking's Euclidean approach. Furthermore, we explained in detail why the Euclidean path integral is divergent and hence cannot be taken to be a fundamental starting point of the theory. 

We showed in simple minisuperspace examples how the Lorentzian path integral reduces to a perfectly convergent (albeit conditionally convergent) integral over the space of fields. It is very plausible that this result extends to include all dynamical modes. In particular, the fluctuations about our homogeneous, isotropic but sometimes complex saddle point solutions, will themselves possess quadratic, complex actions. It is clear that Picard-Lefschetz theory applies rather trivially to this case and {\it always} yields a convergent measure on the space of field fluctuations. Furthermore, field interactions lead to higher powers in the action and hence better and better convergence when the Picard-Lefschetz approach is employed. It is also clear how unitarity is recovered for cosmological backgrounds corresponding to real (or nearly real) saddle point solutions, because the starting point of the whole theory is a path integral over real fields. We believe these arguments, as well as the  examples we have investigated in detail, provide compelling evidence that the Lorentzian formulation of quantum cosmology is to be preferred. 

As we have already mentioned, there is significant overlap between this work and that of \cite{Gielen:2015uaa,Gielen:2016fdb} which describes a quantum cosmological bounce for conformal-invariant matter and free scalar fields, also in a Lorentzian formulation.  In that context, two of us have recently shown how cosmological time can emerge~\cite{Feldbrugge:2017}. 

An obvious extension will be the inclusion of other types of matter, such as  pressure-free matter and scalar fields with nontrivial potentials. In particular it will be interesting to revisit previous uses of instantons,  in both inflationary \cite{Hartle:2008ng} and ekpyrotic \cite{Battarra:2014xoa,Battarra:2014kga} cosmologies, as well as attempts to describe quantum transitions between contraction and expansion in such models \cite{Bramberger:2017cgf}, in the light of Picard-Lefschetz theory. In this context, exactly solvable models such as those of \cite{Garay:1990re,Bars:2012mt} may provide useful insight. As we have seen, for the simplest cosmology with only a cosmological constant we obtained the same result as Vilenkin, in his ``tunneling'' proposal for the wavefunction of the universe. However, since the logic we have employed is quite different, it remains to be seen whether the two approaches will agree for more complex and realistic models. 

More generally, it will be interesting to see how tunneling, and other nonperturbative quantum gravity processes, can be treated in this Lorentzian-Picard-Lefschetz (LPL) framework. 

\acknowledgments

We would like to thank Martin Bojowald, Angelika Fertig, Davide Gaiotto, Jaume Garriga, Gary Gibbons, Steffen Gielen, James Hartle, Thomas Hertog, Kelly Stelle, and Alex Vilenkin for useful discussions and correspondence. Research at Perimeter Institute is supported by the Government of Canada through Industry Canada and by the Province of Ontario through the Ministry of Research and Innovation.

\bibliographystyle{utphys}
\bibliography{PicardLefschetz}

\appendix
\section{The Feynman propagator and the Wheeler-DeWitt equation in a De Sitter universe}\label{Ap:Prop}
In section \ref{section:HH} we showed that the Feynman propagator for a De Sitter minisuperspace model is given by
\begin{align}
G[q_1;q_0]
=& \sqrt{\frac{3\pi i}{2}}\int_0^\infty \frac{\mathrm{d}N}{N^{1/2}}e^{2\pi^2iS_0}\,,
\end{align}
with
\begin{align}
S_0=N^3 \, \frac{\Lambda^2}{36} + N \left( -\frac{\Lambda}{2}(q_0+q_1) +3k \right) +\frac{1}{N}\left( -\frac{3}{4} (q_1-q_0)^2\right)\,,
\end{align}
in our minisuperspace model of gravity with a positive cosmological constant $\Lambda.$ Starting with the propagator, we can derive the Wheeler-deWitt equation by taking derivatives. The partial derivative of $G$ with respect to $q_1$ is
\begin{align}
\frac{\partial G}{\partial q_1}
=& \sqrt{\frac{3\pi i}{2}}\int \frac{\mathrm{d}N}{N^{1/2}} 2\pi^2 i S_{0,q_1} e^{2\pi^2i S_0 }\nonumber\\
=& \sqrt{\frac{3\pi i}{2}}\int \frac{\mathrm{d}N}{N^{1/2}} 2\pi^2 i\left[ -\frac{N}{2}\Lambda - \frac{3}{2N} (q_1-q_0) \right] e^{2\pi^2i S_0}.
\end{align}
The second order partial derivative of $G$ with respect to $q_1$ is
\begin{align}
\frac{\partial^2 G}{\partial q_1^2} 
=& \sqrt{\frac{3\pi i}{2}} \int \frac{\mathrm{d}N}{N^{1/2}} \left[ 2 \pi^2 i S_{0,q_1q_1} - 4 \pi^4 S_{0,q_1}^2 \right] e^{2\pi^2iS_0}\nonumber\\
=& \sqrt{\frac{3\pi i}{2}} \int \frac{\mathrm{d}N}{N^{1/2}} \left[ -4\pi^4 \left(\frac{N}{2}\Lambda + \frac{3}{2N} (q_1-q_0)\right)^2 - \frac{3\pi^2i}{N} \right] e^{2\pi^2iS_0}\,.
\end{align}
The argument of the integral depends on $N$. We would like to remove this dependence by using the properties of the Lefschetz thimbles. From the fundamental theorem of calculus and partial integration we have that
\begin{align}
\left[N^{-\frac{1}{2}}e^{2\pi^2 i S_0}\right]_0^\infty =& \int \mathrm{d}N \frac{\mathrm{d}}{\mathrm{d}N}\left[N^{-\frac{1}{2}}e^{2\pi^2 i S_0}\right] =-\frac{1}{2}\int \frac{\mathrm{d}N}{N^{\frac{3}{2}}} e^{2\pi^2 i S_0} + 2 \pi^2 i \int \frac{\mathrm{d}N}{N^{\frac{1}{2}}} S_{0,N}e^{2\pi^2 i S_0}\,.
\end{align}
Substituting this relation in the second order partial derivative of $G$ with respect to $q_1$ gives
\begin{align}
\frac{\partial^2 G}{\partial q_1^2}=&
\sqrt{\frac{3\pi i}{2}} \left[\int \frac{\mathrm{d}N}{N^{1/2}} \left[ -4\pi^4 \left(\frac{N}{2}\Lambda + \frac{3}{2N} (q_1-q_0)\right)^2  \right] e^{2\pi^2iS_0} 
- 3\pi^2i \int \frac{\mathrm{d}N}{N^{\frac{3}{2}}} e^{2\pi^2iS_0}\right]\nonumber\\
=&
\sqrt{\frac{3\pi i}{2}} \left[\int \frac{\mathrm{d}N}{N^{1/2}} \left[ -12\pi^4 \left(  \Lambda q_1 - 3k 
\right)  \right] e^{2\pi^2iS_0} + 6 \pi^2 i  \left[ N^{-\frac{1}{2}} e^{2\pi^2 i S_0}\right]_0^\infty
\right]\nonumber\\
=& -12\pi^4 \left(  \Lambda q_1 - 3k\right) G + 6 \pi^2 i  \sqrt{\frac{3\pi i}{2}}\left[ N^{-\frac{1}{2}} e^{2\pi^2 i S_0}\right]_0^\infty.
\end{align}
Hence
\begin{align}
\frac{\partial^2 G}{\partial q_1^2} + 12\pi^4 \left(  \Lambda q_1 - 3k\right) G =& 6 \pi^2 i  \sqrt{\frac{3\pi i}{2}}\left[ N^{-\frac{1}{2}} e^{2\pi^2 i S_0}\right]_0^\infty\,.
\end{align}
The contribution corresponding to the limit $N\to \infty$ vanishes, since the Lefschetz thimble is constructed such that $e^{2\pi^2 i S_0}\to 0$ and $\frac{1}{\sqrt{N}}$ certainly becomes small in this limit. The propagator $G$ thus satisfies
\begin{align}
\frac{\partial^2 G}{\partial q_1^2} + 12\pi^4 \left(  \Lambda q_1 - 3k\right) G 
=& - 6 \pi^2 i  \sqrt{\frac{3\pi i}{2}} \lim_{N\to 0} \frac{e^{2\pi^2 i S_0}}{\sqrt{N}}\,.
\end{align}
In the limit $N \to 0$ the action diverges as 
\begin{align}
S_0 \to \frac{1}{N}\left( -\frac{3}{4} (q_1-q_0)^2\right).
\end{align}
Writing $N=in$, since the Lefschetz thimbles approach the origin along the imaginary axis, 
\begin{align}
\lim_{N\to 0} \frac{e^{2\pi^2 i S_0}}{\sqrt{N}} =& \sqrt{\frac{2\pi}{i}} \lim_{n \to 0} \frac{ e^{-3\pi^2  \frac{(q_1-q_0)^2}{2n}}}{\sqrt{2\pi n}}
=  \sqrt{\frac{2}{3 \pi i}} \delta(q_0-q_1)\,.
\end{align}
So, reinstating $\hbar,$ the Wheeler-deWitt (propagator) equation is given by
\begin{align} \label{WdW}
\hbar^2 \frac{\partial^2 G}{\partial q_1^2} + 12\pi^4 \left(  \Lambda q_1 - 3k\right) G 
=&  - 6 \pi^2 i  \delta(q_0-q_1).
\end{align}
Note that, had we integrated over a contour from $N=-\infty$ to $N=+\infty$ (ignoring the singularity at $N=0$) the Dirac delta function term on the right hand side would have been absent.
\end{document}